\documentclass[review]{elsarticle}

\usepackage[T1]{fontenc}
\usepackage[utf8]{inputenc}

\usepackage[colorlinks=true,allcolors=blue]{hyperref}
\usepackage{amsmath,amssymb,amsfonts}
\usepackage{graphicx}
\usepackage{xcolor}
\usepackage[all]{nowidow}
\usepackage[inline]{enumitem}
\usepackage[scaled]{beramono}
\usepackage{booktabs} 
\usepackage[colorinlistoftodos]{todonotes}
\usepackage[tight,footnotesize]{subfigure}
\usepackage{multirow}
\usepackage{balance}
\usepackage{listings}
\usepackage{tcolorbox}
\usepackage{cleveref}
\usepackage{soul}
\usepackage{lscape}
\usepackage{pifont}
\usepackage{xcolor,colortbl}
\usepackage{pdflscape}

\usepackage{algorithm}
\usepackage{algpseudocode}

\definecolor{mygreen}{rgb}{0,0.6,0}
\definecolor{mygray}{rgb}{0.95,0.95,0.95}
\definecolor{myred}{rgb}{0.5,0,0}

\lstdefinestyle{JavaStyle} {
  backgroundcolor=\color{white},   
  commentstyle=\color{mygreen}, 
  breakatwhitespace=false,
  keywordstyle=\color{violet},
  language=Java,
  stringstyle=\color{blue},
  basicstyle=\scriptsize\ttfamily,
  showstringspaces=false }

\lstdefinestyle{SOStyle} {
  backgroundcolor=\color{mygray},   
  commentstyle=\color{mygray}, 
  breakatwhitespace=false,
  keywordstyle=\color{blue},
  language=Java, 
  stringstyle=\color{myred},
  basicstyle=\scriptsize,
  frame=single,
  showstringspaces=false }

\lstdefinestyle{NormalStyle} {
	backgroundcolor=\color{mygray},   
	commentstyle=\color{mygray}, 
	breakatwhitespace=false,
	keywordstyle=\color{blue},
	language=Java, 
	stringstyle=\color{myred},
	basicstyle=\small\ttfamily,
	showstringspaces=false }

\usepackage{xspace}
\usepackage{amssymb}
\usepackage[colorinlistoftodos]{todonotes}
\usepackage{xcolor}

\newcommand*{\ie}{i.e.,\@\xspace}
\newcommand*{\eg}{e.g.,\@\xspace}

\newcommand*{\PR}{PisaRec\@\xspace}

\newcommand\rev[1]{\textcolor{black}{#1}}

\makeatletter
\newcommand*{\etc}{%
	\@ifnextchar{.}%
	{etc}%
	{etc.\@\xspace}%
}
\makeatother
\newcommand*{\etal}{\emph{et~al.}\@\xspace}

\newcommand*\circled[1]{\tikz[baseline=(char.base)]{\color{black} 
		\node[shape=circle,draw=cyan,fill=black!10!white,inner sep=.3pt] (char) 
		{{{\texttt\textbf #1}}};}}

\definecolor{darkgray}{gray}{0.78}
\definecolor{lightgray}{gray}{0.85}
\definecolor{verylightgray}{gray}{0.95}
\definecolor{codegreen}{rgb}{0,0.6,0}
\definecolor{codegray}{rgb}{0.5,0.5,0.5}
\definecolor{codepurple}{rgb}{0.58,0,0.82}
\definecolor{backcolour}{rgb}{0.95,0.95,0.92}


\lstdefinestyle{java}{
	backgroundcolor=\color{backcolour},   commentstyle=\color{codegreen},
	keywordstyle=\color{magenta},
	numberstyle=\tiny\color{codegray},
	stringstyle=\color{codepurple},
	basicstyle=\ttfamily\scriptsize,
	breakatwhitespace=false,         
	breaklines=true,                 
	captionpos=b,                    
	keepspaces=true,                 
	numbers=left,                    
	numbersep=5pt,                  
	showspaces=false,                
	showstringspaces=false,
	showtabs=false,                  
	tabsize=2
}

\newcommand{\rqfirst}{\textbf{RQ$_1$}: \emph{How well does the users’ self-assessment reflect their privacy category?}} 

\newcommand{\rqsecond}{\textbf{RQ$_2$}: \emph{Which sets of questions are relevant for assessing privacy concerns?}} 

\newcommand{\rqthird}{\textbf{RQ$_3$}: \emph{To which extent is \PR able to utilize the obtained categorization in recommending relevant privacy settings to users?}} 

\begin{document}
	\begin{frontmatter}
	


	\title{Leveraging Privacy Profiles to Empower Users in the Digital Society}

	\author[univaq]{Davide Di Ruscio}\corref{cor1}	
	\author[univaq]{Paola Inverardi}
	\author[univaq]{\\Patrizio Migliarini}
	\author[univaq]{Phuong T. Nguyen}
	
	\address[univaq]{Universit\`a degli studi dell'Aquila	L'Aquila, Italy\\
		\{davide.diruscio, paola.inverardi, patrizio.migliarini, phuong.nguyen\}@univaq.it}
	\cortext[cor1]{Corresponding author}
	
	\begin{abstract}
		Privacy and ethics of citizens are at the core of the concerns raised by our increasingly digital society. Profiling users is standard practice for software applications triggering the need for users, also enforced by laws, to properly manage privacy settings.  Users need to manage software privacy settings properly to protect personally identifiable information and express personal ethical preferences. AI technologies that empower users in their interaction with the digital world by reflecting their personal ethical preferences can be key enablers of a trustworthy digital society. This paper focuses on the privacy dimension and contributes a step in the above direction through an empirical study on an existing dataset collected from the fitness domain. 
		The study aims to understand which set of questions and settings is more appropriate to differentiate users according to their privacy preferences. The experimental results reveal that a compact set of semantic-driven questions (about general privacy preferences) helps distinguish users better than a complex domain-dependent one (concerning the fitness domain). Based on the study outcome, we design and implement a recommender system to provide users with suitable recommendations with respect to privacy choices. 
		We then show that the proposed recommender system provides relevant settings to users, obtaining high prediction accuracy. 		
	\end{abstract}


	\end{frontmatter}
	%
	
		\section{Introduction}
	\label{sec:Introduction}

Privacy and ethics of citizens are at the core of the concerns raised by our increasingly digital society. Profiling users is standard practice for software applications triggering the need for users, also enforced by laws, to properly manage privacy settings and moral preferences. This deals with the way users give their consent to storing, sharing to third parties, as well as disseminating sensitive personal information and express moral preferences like, for example, ticking to pay a \emph{decarbonization tax}. 
\rev{Mobile apps have been becoming increasingly popular as they can provide users with a wide range of functionalities. For different reasons, apps often require access to intimate information about the users and hosting device, triggering privacy concerns. This requires proper management of privacy, with the ultimate aim of protecting users' preferences as well as personally identifiable information.}

In this paper, we focus on the privacy dimension of an ordinary user with little technical knowledge of the privacy mechanisms of the digital systems she seamlessly uses but with an evident moral character. While choosing strict settings may help protect her data, this may prevent the complete availability of the functionalities provided by the software. In contrast, loosening privacy settings mitigates the restriction on functionalities, but it may come with the price of compromising her data privacy. In this respect, Artificial Intelligence (AI) technologies can empower the user in maintaining a reasonable trade-off between accessibility and protection, and reflecting the user privacy preferences can be the key enabler of a trustworthy digital society.

Understanding the commonalities and differences among users based on profiles has been among the main issues in data privacy research~\cite{westin_biblio,Kumaraguru_Cranor:2005,ZHAO2019449}. Categorizing profiles contributes to better identification of users' behaviors and supports administrators in comprehending privacy choices. At the same time, personal profiles may enable the design of functionalities that help users set privacy preferences of the digital technologies they use. Various proposals to categorize or group end-users into clusters based on their security or privacy attitudes/behaviors in specific domains have been made~\citep{10.1007/978-3-540-68825-9_22,7845392}. 
Users' preferences were analyzed in an extensive study~\citep{lin2014modeling} on permission settings from real Android mobile users to recommend personalized default settings. Sanchez \etal~\citep{sanchez_recommendation_2020} analyzed user-privacy preferences in the fitness domain employing a specifically designed questionnaire consisting of both domain-specific and general questions to recommend personalized privacy settings for the fitness apps.

Though a lot of achievements have been reported, as discussed in~\citep{Liu2020_PHD}, we believe that there is still the need to understand how to characterize user's privacy behavior in a general setting. Indeed privacy is a dimension of ethics and should be part of the ethical profile of a user and driven by ethical consideration rather than by contextual attitudes or practices in given domains. For example, relying on the analysis of current or past users' preference settings as in~\citep{lin2014modeling} does not guarantee a correspondence between what users believe as their general privacy profile and what they actually (can) do when setting privacy preferences. Moreover, data privacy awareness in the digital society is only recently exiting the specialists' fields (legal, ethical, economic, social) to impact the wider society. The pandemic has also dramatically advanced the penetration of digital technologies in the society from market to education \citep{McKinsey2020,/content/publication/bb167041-en}. This means that a large body of collected data on privacy settings may not reflect the attitude and attention to privacy that present users have and will have in the future.

In this work, we explore a different research direction by relying on the data of the study in the fitness domain \citep{sanchez_recommendation_2020}  that were \rev{collected by means of a questionnaire and a simulator.}\footnote{We thank Prof. Dr. Ilaria Torre, University of Genoa (Italy) for providing us with the privacy dataset~\citep{sanchez_recommendation_2020}.} We analyze both general and domain-specific questions with the aim of  \emph{(i)} identifying general questions that reflect moral attitudes of the users;
and \emph{(ii)} recommending privacy preferences accordingly. 
\rev{Moreover, we design and implement a recommender system~\cite{PUJAHARI2019126} to provide users with suitable recommendations with respect to privacy choices. The experimental results are positively interesting, revealing that a compact set of general questions helps distinguish users better than a more complex domain-dependent one. We also show that the proposed recommender system provides relevant settings to users, obtaining high prediction accuracy.}


The main contributions of our work are summarized as follows.
\begin{itemize}
	\item We investigate which sets of (general) privacy questions are more relevant for classifying users with respect to their privacy moral preferences.
	\item By means of an empirical evaluation, we show that self-assessment about privacy attitudes given by users does not reflect the way they act in practice.
	\item We develop \PR, a recommender system to provide suitable privacy settings that reflect user preferences. \rev{This aims to help users relieve the burden of setting privacy configurations when they go online.}
\end{itemize}

\rev{We organize the paper into the following sections. In Section~\ref{sec:Background}, we present a motivating example and a categorization of privacy profiles. Section~\ref{sec:Approach} describes the proposed approach which makes use of both unsupervised and supervised learning to handle user profiles. The methods used to evaluate our approach are detailed in Section~\ref{sec:Evaluation}. We report and analyze the experimental results in Section~\ref{sec:Results}. Discussion related to the limitations and threats to validity are provided in Section~\ref{sec:Discussion}. We review related work in Section~\ref{sec:RelatedWork}. Finally, Section~\ref{sec:Conclusions} sketches future work and concludes the paper.}


	\section{Background}	
	\label{sec:Background}

The following example illustrates the need for personalized automated privacy assistance that a user interacting with multiple systems at a time may require.  Then we briefly report the most relevant aspects for our research taxonomies for privacy profiles proposed in the literature. 

\subsection{Motivating example}




After a long day at work, Alice is at the subway station. After the pandemic outbreak, she will meet pals at the cinema. She is on time but learns that she cannot buy a ticket from the subway station attendant due to rigorous hygiene regulations. In addition, vending machines are out of commission for contact-less technology upgrades. Instead, a QR code and simple instructions to buy an electronic ticket online are posted in front of the vending machines. 
Her train is about to arrive, she opens her camera app and frames the QR code. The site structure appears in a split second, but as Alice scrolls down to find the ticket she needs, a popup asks for her privacy settings. Above a very long list of radial button options about disclosing GPS position, information about her mobile phone, consent to save various types of cookies on her device, share her list of contacts, etc., she is presented with three buttons: \emph{accept all}, \emph{strictly necessary}, \emph{decline all}. 

Alice is very concerned about her privacy, and when not strictly necessary for the purpose she wants to perform, she does not wish to disclose private information. Since the service she is asking for is simple as asking for a one-ride ticket, she clicks \emph{decline all}. The next page seems to load slowly, images and structure are shown in a non-adaptive way, so she has to pinch-in to zoom and scroll to read the text that informs her that a cryptographic key used for her session management cannot be stored due to her preferences, so the session is not secure also the page asks her to choose language, timezone, type of device and the web browser she is using,  payment options, etc. While reading, Alice realizes that her train is about to arrive at the station. So, she decides to click the back button on her browser, reload the page and click \emph{strictly necessary} when prompted. The site then stays fast and steady, adapted to the display of her device, prompting if she wants to take a one-ride ticket or a full day one. Her mobile wallet handles the payment instantly, and she receives her ticket just before the train comes. On time to the cinema, Alice enjoys the film with her friends, soon forgetting the online ticket purchase experience. Her preferences are saved on her phone, so she will buy train tickets quickly and easily in the future.
Alice does not know that the \emph{strictly necessary} option, although excluding third-party tracking and marketing, includes all alternatives that are strictly essential to all services offered by the booking site, including -- the lower price inter-city ticket that requires GPS tracking, the discounted price for kids that requires age disclosure, discount for army and state officials who must check other installed mobile applications, as well as the train pass app to see if the ticket is part of a booklet, etc.

Analogously to various studies 
notably Liu \etal \citep{liu_follow_2016}, we believe that a software technology should assist Alice in automatically selecting the options that, on the one hand, are needed for what she wants to do and, on the other hand, are compliant with her moral preferences.

In this work, we show that it is possible to protect users by first understanding their privacy profiles, which can be automatically identified by considering a small set of general and domain-independent questions that are shown to be enough to reflect the user's moral attitude.
Thus, our approach is to categorize personal privacy profiles from an ethical perspective~\citep{AutiliRIPT19}. Profiles can then be used to automate app and web settings, leveraging recommender systems like in this paper or other technologies. 

\subsection{Categorizations of privacy profiles} \label{sec:Categorization}



\begin{table}
	\center
	\tiny
	\caption{Privacy categories according to different taxonomies (Listed in chronological order).}
	\label{tab:Categories}
	\begin{tabular}{|p{1.7cm}|p{1.5cm}|p{1cm}|p{1cm}|p{1.1cm}|p{1.1cm}|p{1.7cm}|} \hline
		
		
		\textbf{How important is privacy to you?} & \cellcolor{green!20}NOTHING & 
		\multicolumn{2}{c|}{
			\cellcolor{yellow!30}LITTLE} & 
		\multicolumn{2}{c|}{
			\cellcolor{pink!50}QUITE}   & 
		\cellcolor{pink!90}VERY \\ \hline
		
		\textbf{Segmentation}~\cite{westin_biblio} & 
		\cellcolor{green!20}Unconcerned & 
		\multicolumn{4}{c|}{
			\cellcolor{yellow!60}Pragmatists}   & 
		\cellcolor{pink!90}Fundamentalist \\ \hline
		
		\textbf{Privacy Personas}~\cite{10.1145/2858036.2858214} &
		\cellcolor{green!20}Marginally concerned &
		\multicolumn{2}{c|}{
			\cellcolor{yellow!40}Amateurs}  &
		\cellcolor{pink!20}Technicians &
		\cellcolor{pink!40}Lazy Experts &
		\cellcolor{pink!90}Fundamentalists \\ \hline

		\textbf{Philosophies}~\cite{10.1093/jamia/ocz010} & 
		\cellcolor{green!20}Fatalism & 
		\cellcolor{yellow!30}Nothing to hide &
		\cellcolor{yellow!50}Something to hide &
		\cellcolor{pink!20}Trade-off &
		\cellcolor{pink!40}Personal Resp. &
		\cellcolor{pink!90}Moral right \\ \hline

		\textbf{Privacy Clustering}~\cite{sanchez_recommendation_2020} &
		\cellcolor{green!20}Unconcerned &
		\multicolumn{2}{c|}{
			\cellcolor{yellow!40}Socially active}   &
		\cellcolor{pink!20}Health-focused & 
		\cellcolor{pink!40}Minimal & 
		\cellcolor{pink!90}Anonymous (Strict) \\ \hline
		
		\textbf{Self-assessment}~\cite{sanchez_recommendation_2020} & 
		\cellcolor{green!20}Conservative & 
		\multicolumn{2}{c|}{
			\cellcolor{yellow!40}Unconcerned} & 
		\multicolumn{2}{c|}{
			\cellcolor{pink!40}Fence-Sitter}   & 
		\cellcolor{pink!90}Advanced Users \\ \hline
		
		\textbf{Our proposed categorization} & 
		\cellcolor{green!20}INATTENTIVE & 
		\multicolumn{4}{c|}{
			\cellcolor{yellow!60}INVOLVED/ATTENTIVE}   & 
		\cellcolor{pink!90}SOLICITOUS \\ \hline
		
	\end{tabular}
\end{table}

Table~\ref{tab:Categories} gives a summary of the most notable taxonomies of privacy categories. Starting from the question: ``\emph{How important is privacy to you?}'' 
from left to right of the table, 
an increasing level of privacy concerns is shown.
Westin~\citep{westin_biblio} proposed the first categorization of user profiles with three levels, \ie \emph{Unconcerned}, \emph{Pragmatists}, and \emph{Fundamentalist}. Since then, there have been other studies that follow up and develop this initial taxonomy. In particular, 
Dupre \etal~\citep{10.1145/2858036.2858214} expanded it proposing five categories:
\emph{Marginally concerned}, \emph{Amateurs}, \emph{Technicians}, \emph{Lazy Experts}, and \emph{Moral right}.
Schairer \etal~\citep{10.1093/jamia/ocz010} came up with even more, \ie six categories, where the answer \emph{Little} is split into \emph{Nothing to hide}, and \emph{Something to hide}; and \emph{Quite} is made of \emph{Trade-off}, and \emph{Personal Resp}. 
Recently, Sanchez \etal~\citep{sanchez_recommendation_2020} proposed a more compact categorization, where users are grouped into four categories, \emph{Privacy concerns}, \emph{Unconcerns}, \emph{Fence-Sitter}, and \emph{Advanced Users}.
As it appears from the table, category refinement happens in the middle category and may depend on the application domain as well as on the amount of input data. Further categories are contextual and can be obtained as specialization of personal profiles based on the single user's experience. Therefore, as starting point three categories provide three clearly distinguishable moral  
attitudes.   



The categorization we propose names the three clusters as \emph{Inattentive}, \emph{Attentive}, and \emph{Solicitous}. While \emph{Inattentive} means that users do not care about privacy, \emph{Solicitous} corresponds to an opposite attitude, where users are completely aware of privacy issues. 
The \emph{Attentive} category is something in between, and covers both \emph{Little} and \emph{Quite} answers to the question ``\emph{How important is privacy to you?}''







	\section{Proposed Approach}	
	\label{sec:Approach}

\begin{figure*}[t!]
	\centering    
	\includegraphics[width=0.99\textwidth]{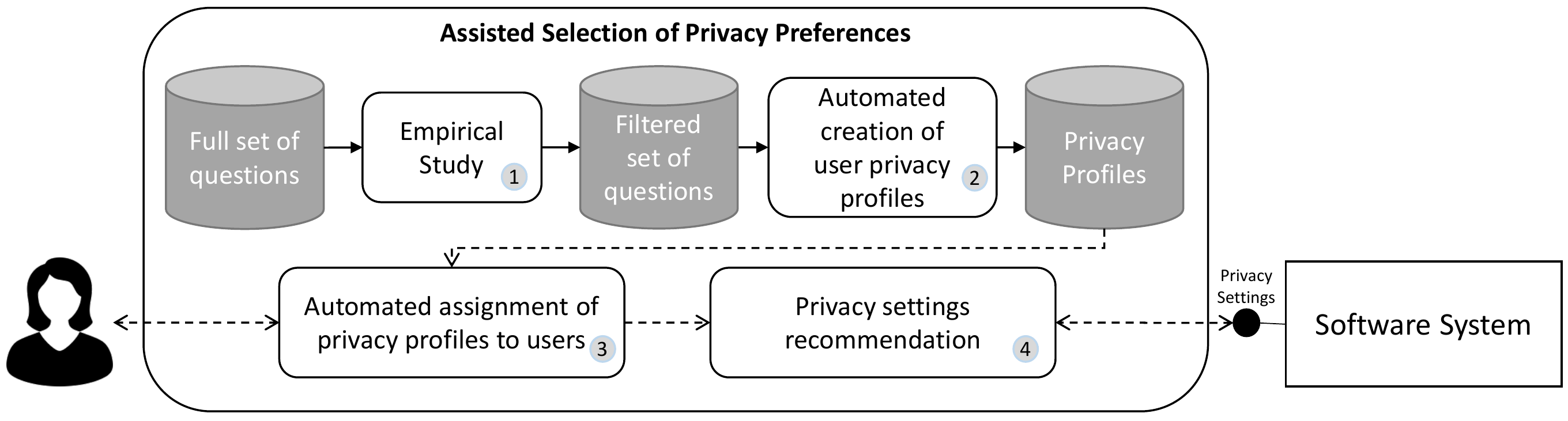}
	\caption{The proposed approach.}
	\label{fig:Assisted}
\end{figure*}

Typically, users specify privacy preferences by directly interacting with the privacy settings provided by the used software. Similar to other techniques \citep{AutiliRIPT19,liu_follow_2016,Liu2020_PHD} we propose an approach that relies on a software layer that automatically identifies privacy profiles and interacts with the user or the software system to recommend privacy preferences accordingly.  

Concerning what we present in this paper, the assisted selection phase of privacy preferences started on training data consisting of general, domain-specific, and app-specific answers given to the questions defined in \citep{sanchez_recommendation_2020} (see \textit{``Full set of  questions''} in Figure \ref{fig:Assisted}). 
We have empirically analyzed the full set of questions to identify the corresponding subset (consisting of general questions) that is sufficient to automatically identify our three user privacy profile categories, i.e., \emph{Inattentive}, \emph{Attentive}, and \emph{Solicitous} (see activity \circled{1} in Figure~\ref{fig:Assisted}).

The automated creation of user privacy profiles phase (see activity \circled{2} in Figure~\ref{fig:Assisted}) relies on an unsupervised clustering module, which can automatically group users in the training data. 
The	automated assignment of privacy profiles to users phase (see \circled{3}) relies on a supervised classifier using a feed-forward neural network to automatically assign to the given user the corresponding privacy profile among one of those identified in \circled{2}.  
Finally, a recommender system is used to further validate the activities \circled{1} and \circled{2}, and to provide users with privacy settings recommendations (see activity \circled{4}) according to the privacy settings of other users belonging to privacy profiles as detected in \circled{3}.

Details on the results obtained from the performed \texttt{Empirical Study} are given in the  \textit{Experimental results} section, whereas the activities \circled{2}, \circled{3}, and \circled{4} are described as follows.

\subsection{Automated creation of user privacy profiles} \label{sec:Clustering}

\begin{figure}[h!]
	\centering
	\includegraphics[width=0.68\textwidth]{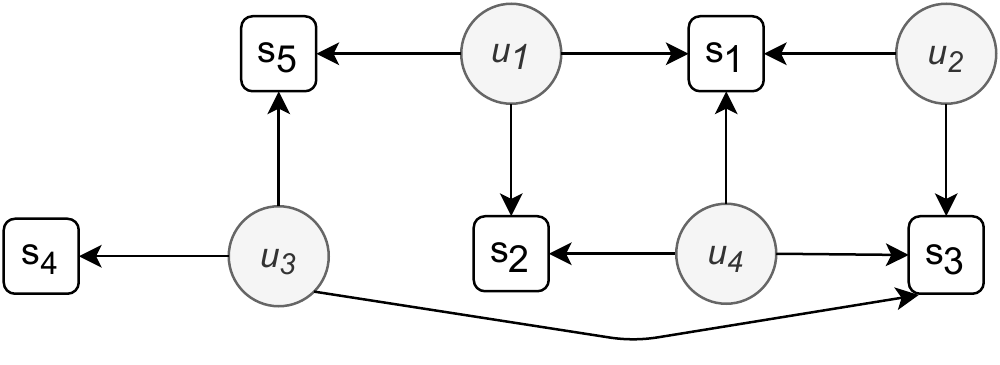}
	\caption{Graph representation of users and privacy settings.}
	\label{fig:Graph}
\end{figure}

To automatically create user privacy profiles, we employed a clustering process by relying on 
%
%
the graph-based representation of users and privacy settings as shown in Figure~\ref{fig:Graph}. This representation is also used by the developed neural network for classifying users presented in Section~\ref{sec:NeuralNetwork}.
%


\rev{Each user $u$ is represented by a vector $\phi=(\phi_{1},\phi_{2},..,\phi_{F})$, where $\phi_{i}$ is the weight of term $s_{i}$, computed as the \emph{term-frequency inverse document frequency} value as follows: }

\begin{equation}\label{eqn:TFIDF}
\phi_{i} = f_{s_{i}} \times log(\frac{ \left | P \right |}{a_{s_{i}}})
\end{equation}



The similarity between two users $u$ and $v$ is computed using their corresponding feature vectors $\phi=\{\phi_{i}\}_{i=1,..,F}$ and $\omega=\{\omega_{j}\}_{j=1,..,F}$ by means of the cosine similarity function: 

\begin{equation} \label{eqn:CosineSimilarity}
sim(u,v)=\frac{\sum_{t=1}^{n}\phi_{t}\times \omega_{t}}{\sqrt{\sum_{t=1}^{n}(\phi_{t})^{2} }\times \sqrt{\sum_{t=1}^{n}(\omega_{t})^{2}}} 
\end{equation}

where $n$ is the cardinality of all settings that were set to $1$ by both $u$ and $v$. 
Intuitively, $u$ and $v$ are characterized by using vectors in an $n$-dimensional space, and Equation~\ref{eqn:CosineSimilarity} measures the cosine of the angle between them. \rev{As an example, in Fig.~\ref{fig:Graph}, we see that the two users $u_2$ and $u_4$ are similar since they both set two settings $s_1$ and $s_3$.} 

A set of $n$ users is grouped into $\kappa$ pre-defined number of clusters, with the aim of maximizing both the similarity among instances within a single cluster, and the dissimilarity among 
independent clusters. To this end, we calculate the distance between every pair of users and feed as the input for the clustering engine. The K-medoids algorithm~\citep{PARK20093336} has been chosen to group users into clusters due to its simplicity and efficiency.


In the clustering process, the distance scores, computed as $d_{C}(u,v) = 1- sim_{C}(u,v)$, are used to assign users to clusters. Initially, a set of medoids (users) is generated randomly, then a medoid is selected as the user in the cluster that has minimum average distance to all the other users in the cluster. Afterwards, users are assigned to the cluster with the closest medoid, using a greedy strategy~\citep{PARK20093336}.

\subsection{Automated assignment of privacy profiles to users} \label{sec:NeuralNetwork}

Supervised learning algorithms can simulate humans' learning activities, mining knowledge from labeled data and performing predictions for unknown data~\citep{10.5555/523781}. \rev{Among others, neural networks have been widely adopted in various applications, including pattern recognitions~\citep{Bishop:1995:NNP:525960}, or forecasting~\citep{RePEc:eee:intfor:v:14:y:1998:i:1:p:35-62}. A feed-forward neural network consists of connected layers of neurons, where the output of a layer is transferred to the next layer's neurons, except for the output layer.}
%
%
%

\begin{figure}[h!]
	\centering
	\includegraphics[width=0.68\textwidth]{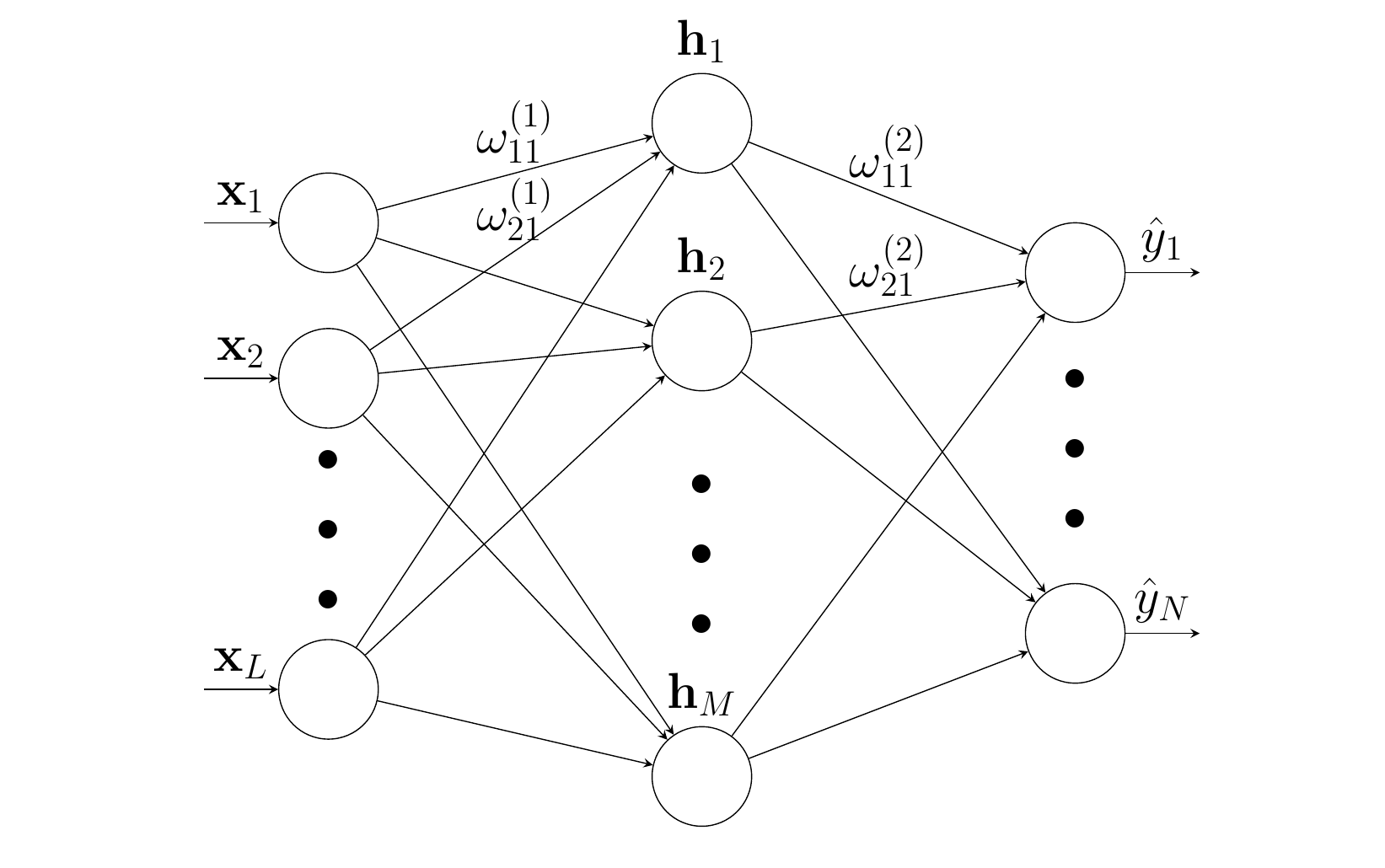}
	\caption{\rev{A three-layer neural network.}}
	\label{fig:NeuralNetwork}
\end{figure}

We built a feed-forward neural network to classify users into different privacy groups, using preferences as features. The network consists of three layers explained as shown in Figure~\ref{fig:NeuralNetwork}. The input layer has $L$ neurons, being equal to the number of input settings, \ie~$X=(x_{1},x_{2},...,x_{L})$. The middle layer consists of $M$ neurons, \ie~$H=(h_{1},h_{2},...,h_{M})$, $M$ can be configured during the evaluation. There are $\kappa$ neurons in the output layer, corresponding to $\kappa$ output categories, \ie~$\hat y=(\hat y_{1},\hat y_{2},..,\hat y_{\kappa})$. The predicted value $\hat y_{k}$ for neuron $k$ of the output layer is computed to minimize the error between the real values and the predicted ones. As discussed in the  \textit{Experimental results} section, the conceived neural network has played an important role in the performed analysis, especially to understand to what extent self-declared privacy profiles reflect the actual user category.

\subsection{Privacy settings recommendation} \label{sec:RecSys}



\begin{figure}[h!]
	\centering
	\includegraphics[width=0.60\textwidth]{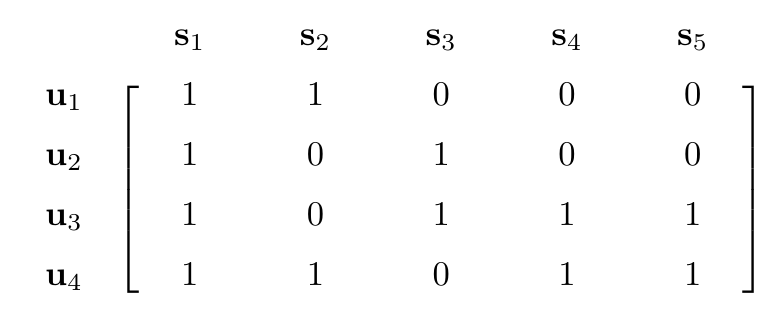}
	\caption{A user-setting matrix.}
	\label{fig:UserItemMatrix}
\end{figure}

We conceptualize \PR, a \textbf{P}r\textbf{i}vacy \textbf{s}ettings \textbf{a}ssistant running on top of a \textbf{Rec}ommender system to provide users with suitable data protection configurations. \PR works based on the assumption that \emph{``if users of the same privacy profile already share some common privacy settings, then they are supposed to share additional similar settings''}~\citep{Schafer:2007:CFR:1768197.1768208}. In this way, we utilize the proposed graph-based representation to model the relationship among users and use a collaborative-filtering algorithm \citep{Aggarwal2016} to recommend missing settings.  
To feed as input for the recommendation engine, we adopt the \emph{user-item} paradigm \citep{DBLP:conf/rweb/NoiaO15}, in which each user corresponds to one row, a column represents each setting. In this way, a cell in the matrix dictates the rating given by a user to a setting.
The two values 0 and 1 correspond to \emph{deny} and \emph{allow}, respectively.
An example of a user-setting matrix for the set of four users and five settings is as follows: $u_1 \ni s_1,s_2$; $u_2 \ni s_1,s_3$; $u_3 \ni s_1,s_3,s_4,s_5$; $u_4 \ni s_1,s_2,s_4,s_5$. Accordingly, the user-item ratings matrix built to model the occurrence of the settings is depicted in Figure~\ref{fig:UserItemMatrix}.

The following collaborative-filtering formula is utilized~\citep{Schafer:2007:CFR:1768197.1768208} to predict the inclusion of a setting $s_{i}$ for user $u$: 

\begin{equation} \label{eqn:Prediction}
r_{u,s_{i}}=\overline{r_{u}}+\frac{\sum_{v \in topsim(u)}(r_{u,s_{i}}-\overline{r_{v}})\cdot sim(u,v) }{\sum_{v \in topsim(u)} sim(u,v) }  
\end{equation}

\noindent
where $\overline{r_{u}}$ and $\overline{r_{v}}$ are the mean of the ratings of $u$ and $v$, respectively; $v$ belongs to the set of \emph{top-k} most similar users to $u$ or neighbour users, \ie $topsim(u)$; $sim(u,v)$ is the similarity between $u$ and a similar user $v$, computed using Equation~\ref{eqn:CosineSimilarity}.

The clusters obtained from the previous section allow us to identify users with similar privacy preferences. Based on the obtained categorization, given an input user, the neural network assigns the user to a specific category. Afterward, we build a graph only for this category following the paradigm in Figure~\ref{fig:Graph}. Such a sub-graph contains fewer nodes and edges than a full graph for all categories, aiming to optimize the computation. On top of this, \PR recommends missing settings to users. The outcome of the computation is a ranked list of probable settings, and we select the top-N of them to present as the final recommendations. 

	\section{Evaluation}	
	\label{sec:Evaluation}

	To study the proposed approach's performance, we first introduce three research questions. Afterward, we describe the dataset and metrics used in our evaluation. 
	

	\subsection{Research questions} \label{sec:ResearchQuestions}
	
	
	\rev{The following research questions are considered to evaluate our proposed approach.}
	
	\begin{itemize}
		\item \rqfirst~As users in the considered dataset \citep{sadeh2021personalized} have been allowed to self-assess their privacy category, we examine if such a self-evaluation 
		reflects their real category. 
		
		\item \rqsecond~We are interested in finding the set of questions that can better distinguish between user profiles. For this research question, we cluster the users with different sets of features, and identify the one that brings the best clustering solution. The aim is to find a set of privacy questions that better represents the user profiles. 
		
		\item \rqthird~We investigate how well the conceived recommender system learns from existing profiles, providing users with additional configurations that reflect their preferences. 
	\end{itemize}

	
	
	
	

	\subsection{Dataset} \label{sec:Dataset}

	
	
	We opted for an existing dataset that has been collected through a domain-specific survey about the usage of a fitness app including user privacy preferences \citep{sanchez_recommendation_2020}.
	%
	As shown in Table~\ref{tab:Datasets}, there are 444 data entries which have been divided into three main groups as follows:
	
	\begin{itemize}
		\item \emph{Domain specific}: This is the set of questions being explicitly related to the fitness activity. There are a total of $202$ questions in this category. 
		\item \emph{App related}: These questions are about the use or setting of the app, consisting of 113 questions. 
		\item \emph{Generic}: This set of questions consists of generic questions that are not related to other groups. There are 129 generic questions in total. 
	\end{itemize}

	
	\begin{table*}[t!]
		\label{tab:questions_summary_table}
		\footnotesize
		\vspace{-.1cm}
		\scriptsize
		\caption{Summary of the dataset.} 
		\centering
		\begin{tabular}{|l|l|p{5.6cm}|c|} 
			\hline
			\textbf{Questions/Data} & \textbf{Alias} & \textbf{Description} & \textbf{\# entries}   \\ \hline 
			\textbf{Domain specific} & D & Questions related to the specific domain (Fitness) & \textbf{202} \\ \hline 
			
			D Subset 1 & DP1 &  Subset of the D set consisting of privacy relevant questions & 123   \\ \hline 
			\textbf{App related} & A & Questions related to the mobile application and the specific software context & \textbf{113}   \\ \hline 
			
			A Subset 1 & AP1 & Subset of the A set consisting of privacy relevant questions & 65   \\ \hline 
			A Subset 2 & AP2 & Subset of the A set that includes only generalizable questions & 6    \\ \hline 
			D + A Subset 0 & S0 &  Privacy related questions from the D and the A sets  (DP1+AP1) & 188    \\ \hline 
			\textbf{Generic} & G & Generic questions not specifically related to the domain (fitness) or the application/software context (mobile app) & \textbf{129}    \\ \hline 
			G Privacy Subset 1 & GP1 & Subset of the G set consisting of privacy relevant questions & 110    \\ \hline 
			G Subset 1 & G1 & Subset of the G set consisting of questions related to the disclosure of information about user's identity with the app & 35    \\ \hline 
			G Data 2 & G2 &  Data concerning the time spent by the users to answer the questionnaire, play with the simulator, and the sum of the two  & 3    \\ \hline 
			G Subset 3 & G3 & Subset of the G set consisting of questions related to the user's identity & 19    \\ \hline 
			G Subset 4 & G4 & Subset of the G set consisting of questions related to the disclosure of private information with the app & 56    \\ \hline 
			G Subset 5 & G5 & Subset of the G set consisting of questions related to the concerns about privacy & 16    \\ \hline 
			
			\textbf{Full dataset} & DATA & Data collected with the questionnaire and the simulator (D+A+G )& \textbf{444} \\ \hline 
		\end{tabular}		
		\label{tab:Datasets}
		\vspace{-.4cm}
	\end{table*}

	\subsection{Evaluation metrics} \label{sec:Metrics}




	\begin{itemize}
		\item \textbf{Compactness}. The metric measures how closely relevant the users within a cluster are~\citep{10.1109/ICDM.2010.35}. In this respect, a lower value represents a better clustering solution and vice versa.
		
		\item \textbf{Silhouette}. It measures how similar a user $u$ is to all the remaining users of the same cluster~\citep{10.1109/ICDM.2010.35}, computed using the following formula: 
		\begin{equation}
		s(u)=\frac{(b(u)-a(u))}{max \{a(u),b(u)\}}
		\end{equation}
		where $a(u)$ is the mean distance between $u$ and the others, $b(u)$ is the minimum mean distance. A silhouette value falls into the range [-1,..+1], where a higher score means a better clustering solution.

		Furthermore, we also use Precision, Recall, ROC curve and AUC to study the performance of the proposed approach.
		
		First, there are the following definitions: True positive (TP) is the settings that match with ground-truth data; False positive (FP) is the recommended settings but do not match with the ground-truth data; False negative (FN): the settings that should be recommended, but they are excluded. Then, the metrics are as follows:
		
		
		\item \textbf{Precision and Recall}. Precision measures the fraction of the number of settings properly classified to the total number of recommended items and 
		Recall (or true positive rate -- TPR) is the ratio of the number of correctly classified items to the total number of items in the ground-truth data. The metrics are defined as follows: 
		
		\vspace{.2cm}
		
		\begin{minipage}{0.50\linewidth}
			\begin{equation} \nonumber 
			P = \frac{TP}{TP+FP}
			\end{equation}
		\end{minipage} 
		\begin{minipage}{0.50\linewidth}
			\begin{equation} \nonumber 
			R = \frac{TP}{TP+FN} = TPR
			\end{equation}
		\end{minipage}
		
		\vspace{.2cm}
		
		
		
		
		\item \textbf{False positive rate (FPR)}. This metric measures the ratio of the number of items that are falsely classified into a category \textbf{c}, to the total number of items that are either correctly not classified, or falsely classified into the category:
		
		\begin{equation}\nonumber
		FPR = \frac{ FP }{TN+FP}
		\end{equation}
		\vspace{.1cm}
		\item \textbf{ROC curve and AUC}. The relationship between FPR and TPR is sketched in a 2D space, using a receiver operating characteristic (ROC) \citep{10.1016/j.patrec.2005.10.010}, which spans from (0,0) to (1,1). An ROC close to the upper left corner represents a better prediction performance. 
		
	\end{itemize}

	\section{Experimental results}	
	\label{sec:Results}

	This section reports and analyzes the experimental results by answering the 
	research questions introduced in Section~\ref{sec:ResearchQuestions}.
	
	
	
	
	\subsection{\rqfirst} 
	
	In the dataset~\citep{sanchez_recommendation_2020} considered in our evaluation, each user has assigned themselves to one of the following four groups: \emph{Privacy Conservative} (Class 0), \emph{Unconcerned} (Class 1), \emph{Fence-Sitter} (Class 2), and \emph{Advanced User} (Class 3). 
	We investigate if the self-assessment is consistent, \ie if all the users properly perceive their real privacy category. This is important since a proper self-clustering can be utilized in additional profiling activities.
	
	\begin{figure}[t!]
		\centering
		\includegraphics[width=0.60\textwidth]{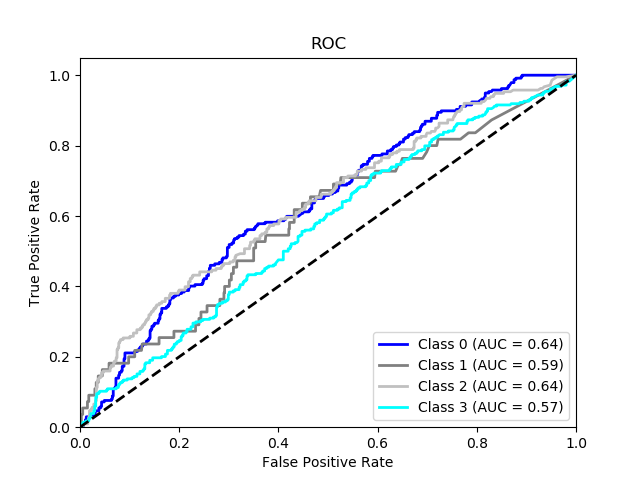}
		\caption{ROC curves with generic questions.}%
		\label{fig:ROC1}
	\end{figure}

	We conducted evaluation using the conceived neural network as the 
	classifier. Such a technique has been successfully applied to classify various types of data, \eg text~\citep{10.1145/3439726}, chemical patterns~\citep{1311506}, metamodels~\citep{8906979}, to name a few. %
	Similarly, we use the privacy settings as features, and the labels specified by humans 
	to train the classifier. We opt for the ten-fold cross validation technique~\citep{Kohavi:1995:SCB:1643031.1643047}, where the dataset is split into ten equal parts, and the evaluation is done in ten rounds.
	The evaluation metrics are computed on the test set, \ie for each user the network predicts a label, which is then compared with the self-assessed label to evaluate the performance. \rev{Finally, ROC curves are sketched by combining the scores obtained from all the ten folds.} 

	\begin{figure}[h!]
		\centering
		\includegraphics[width=0.60\textwidth]{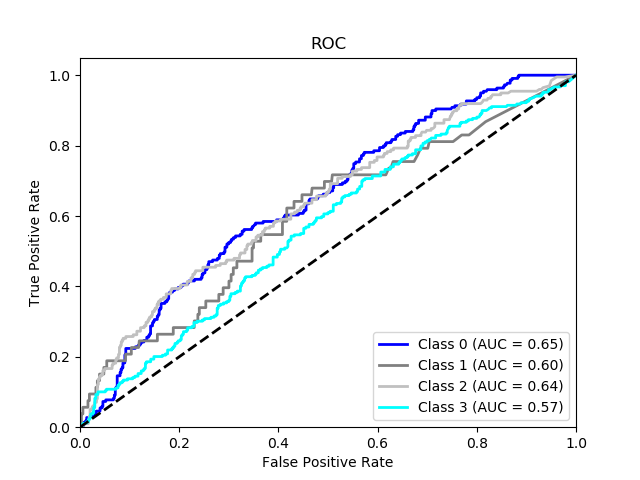}
		\caption{ROC curves with domain specific questions.}
		\label{fig:ROC2}
	\end{figure}

	Figure~\ref{fig:ROC1} and Figure~\ref{fig:ROC2} 
	depict the ROC curves obtained from the classification results for generic and domain specific questions. It is evident that the classifier achieves very low prediction performance on both configurations. In particular, the curves bend over the diagonal line, being close 
	to a random guess. Moreover, the AUC values of the four categories are always lower than 0.65. In other words, we encounter \emph{negative results}, where the neural network fails to predict a proper category for a user. 
	These results suggest that there are noises in the training data~\citep{10.5555/1120115.1120116}, which could possibly be both in the features and the labels.
	
	To confirm the hypothesis, 
	we measure the similarity between each user and all the remaining others. Interestingly, we found out that 96.20\% of the users have very similar users in completely different self-assessed categories.
	This demonstrates that while users share similar 
	preferences, they classify themselves differently, causing a low prediction performance for the neural network.


	%

	\begin{tcolorbox}[boxrule=0.86pt,left=0.3em, right=0.3em,top=0.1em, bottom=0.05em]
		\textbf{Answer to RQ$_{1}$.} 
		The self-assessment given by users does not 
		reflect their real privacy category: Users with highly similar 
		settings perceive themselves as completely different groups. In practice, this means 
		administrators should not rely 
		on such a self-categorization, but have to perform privacy 
		profiling on their own.
	\end{tcolorbox}

	
	%
	

	
	\subsection{\rqsecond} 
	
	As seen in \textbf{RQ$_1$}, the self-assessment given by users is not consistent, thus it is necessary to find another way to group users into clusters. 
	We performed experiments on different subsets of the 
	questionnaire to study the influence of each set on the clustering results. The ultimate aim is to identify a set of questions that helps classify users better. 
	In particular, we are interested in analyzing the following groups of questions:

	
	
	
	\begin{itemize}
		\item \emph{QS$_1$}: It is a set of question sets as follows: Domain specific (\textbf{D}); App related (\textbf{A}); Generic (\textbf{G}) and their combination (\ie \textbf{D}+\textbf{A}+\textbf{G}) named \textbf{COM.}. Furthermore, we also include the set 
		\textbf{G+AP2} where \textbf{AP2} contains  generalizable questions like
		%
		\emph{``Do you believe the company providing this fitness tracker is trustworthy in handling your information?''} Indeed, this is to ask if the company is trustworthy, and therefore we consider it as a general question.
		\emph{QS$_1$} permits to compare compactness and silhouette performances between a single set and combinations of all questions.
		
		\item \emph{QS$_2$}: It is a set of questions sets as follows: \textbf{DP1}, \textbf{AP1}, and \textbf{GP1} that are the subsets of \textbf{D},  \textbf{A}, and \textbf{G} consisting only of privacy relevant questions, respectively. \textbf{COM.} is the union of the three subsets, \ie \textbf{COM. = DP1+AP1+GP1}.  \emph{QS$_2$} permits to understand the actual influence of the privacy-related questions.
		
		
		\item \emph{QS$_3$}: It consists of subsets of generic questions \textbf{G} defined as follows:
		\textbf{G1} are the questions related to disclosure of information about user's identity with the app; \textbf{G2} the questions related to the  time spent by the user in completing the survey; \textbf{G3} the questions related to  user's identity; \textbf{G4} the  questions related to disclosure of private information with the app; \textbf{G5} the questions related to concerns about privacy. \textbf{COM.} is the combination of all the subsets, \ie \textbf{COM. = G1+G2+G3+G4+G5}. \emph{QS$_3$} is to ascertain the influence of the generic questions with respect to the overall set of questions.
		
	\end{itemize}

	\begin{figure*}[t!]
		\centering
		\begin{tabular}{c c }	
			\subfigure[Compactness for QS$_1$]{\label{fig:CompactnessDAG}
				\includegraphics[width=0.46\linewidth]{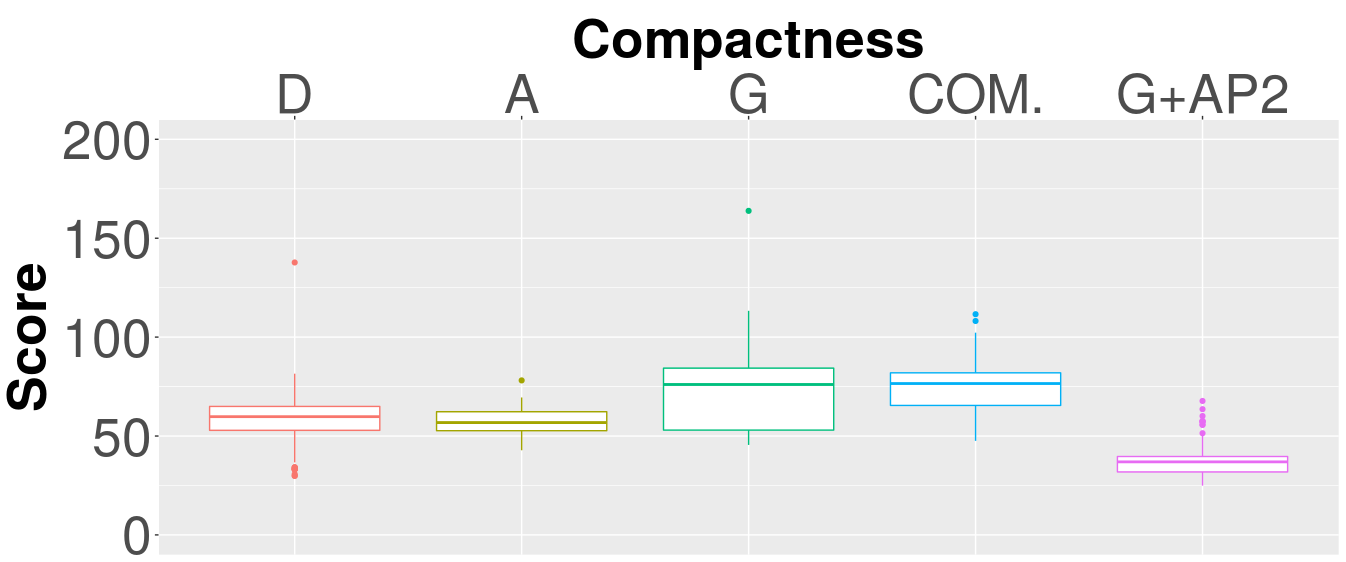}} &
			\subfigure[Silhouette for QS$_1$]{\label{fig:SilhouetteDAG}
				\includegraphics[width=0.46\linewidth]{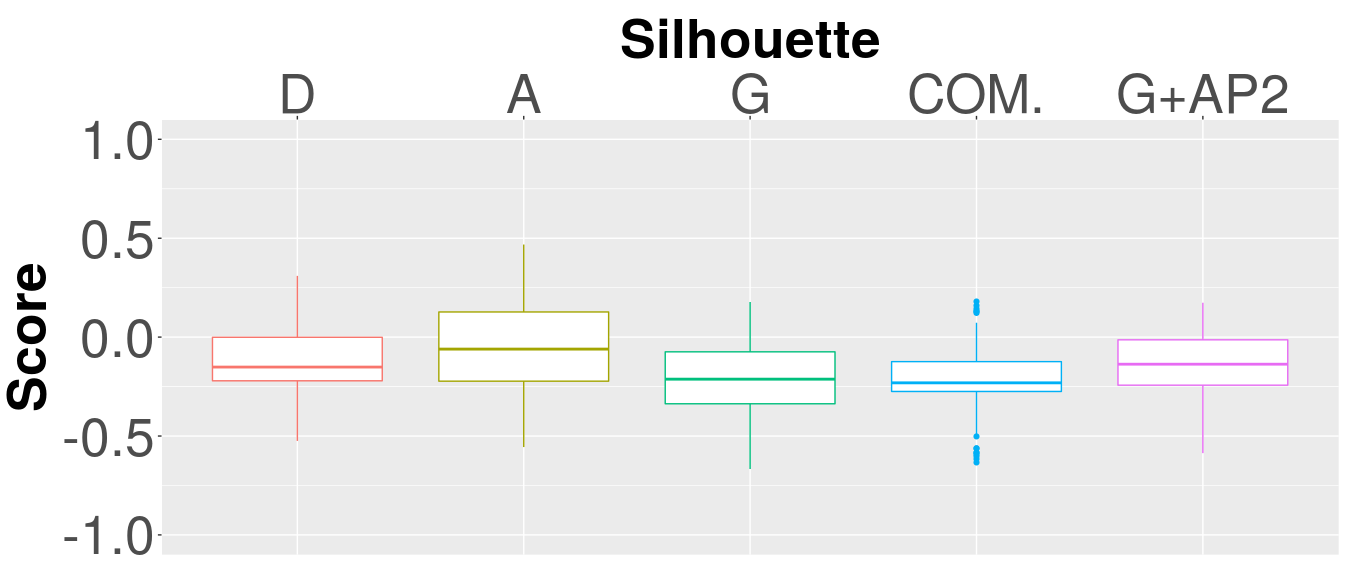}} \\
			
			\subfigure[Compactness for QS$_2$]{\label{fig:CompactnessDP1AP1GP1}
				\includegraphics[width=0.46\linewidth]{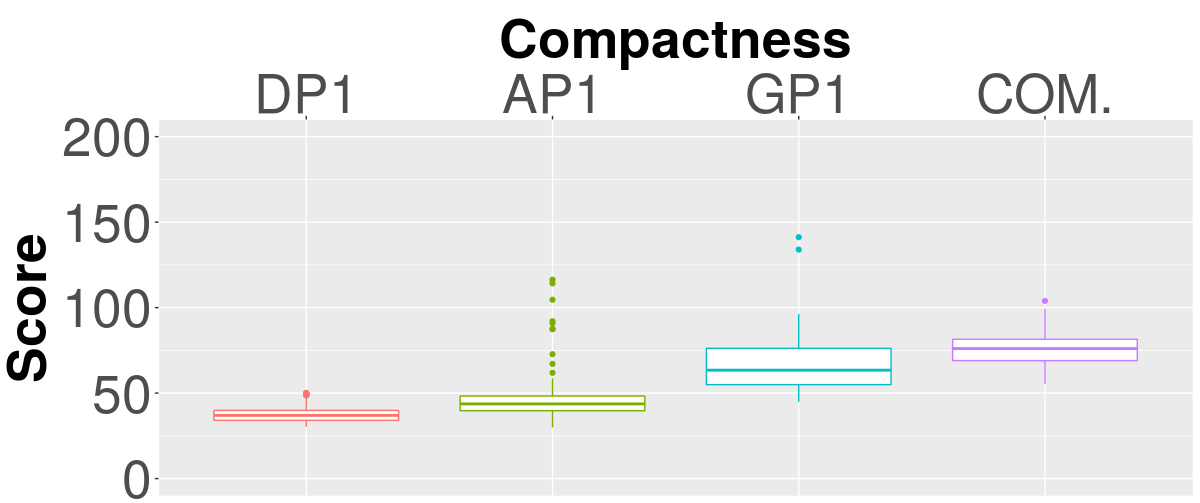}} &
			\subfigure[Silhouette for QS$_2$.]{\label{fig:SilhouetteDP1AP1GP1}
				\includegraphics[width=0.46\linewidth]{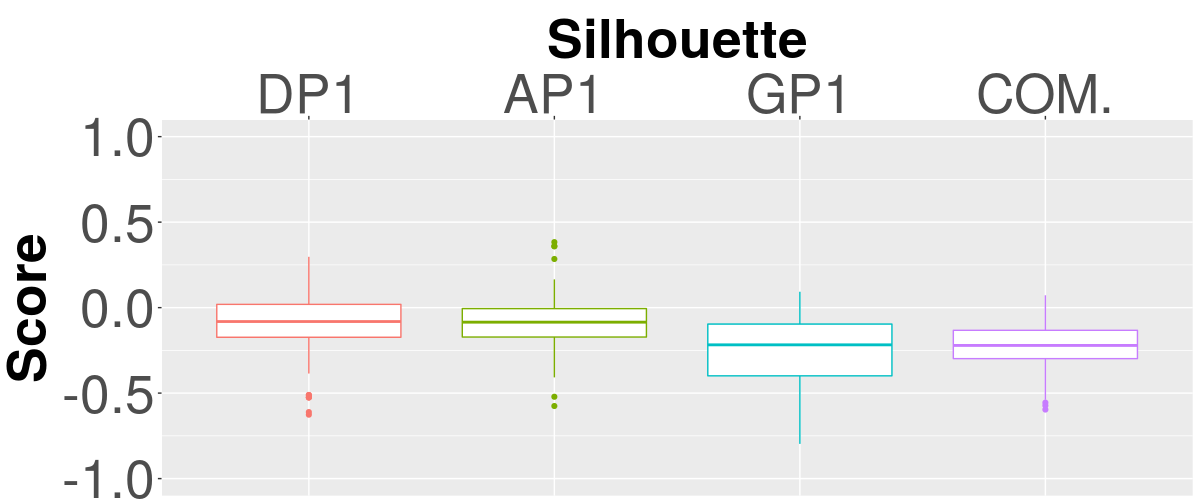}} \\
			
			\subfigure[Compactness for QS$_3$]{\label{fig:CompactnessG1G2G3G4G5}
				\includegraphics[width=0.46\linewidth]{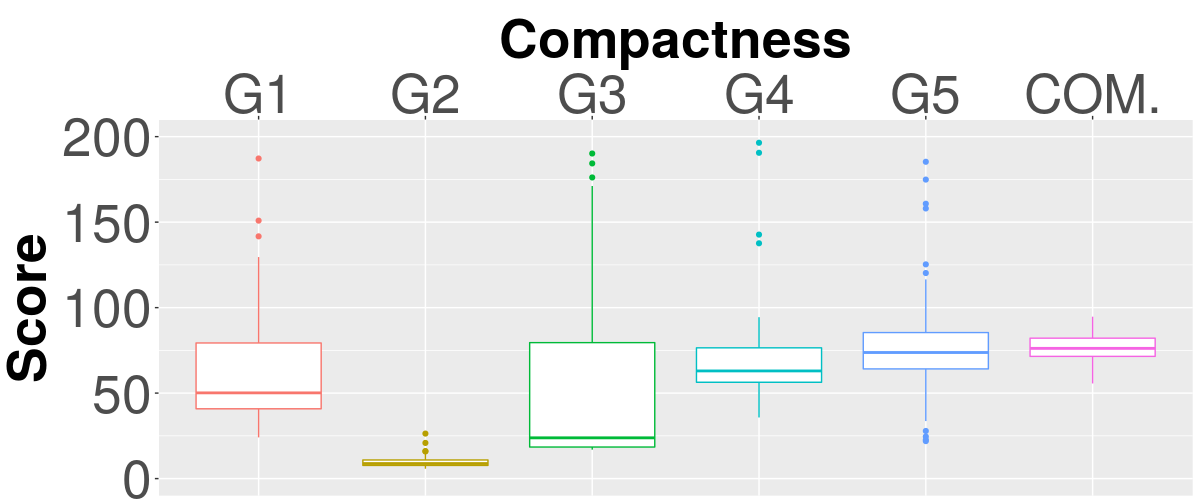}} &
			\subfigure[Silhouette for QS$_3$.]{\label{fig:SilhouetteG1G2G3G4G5}
				\includegraphics[width=0.46\linewidth]{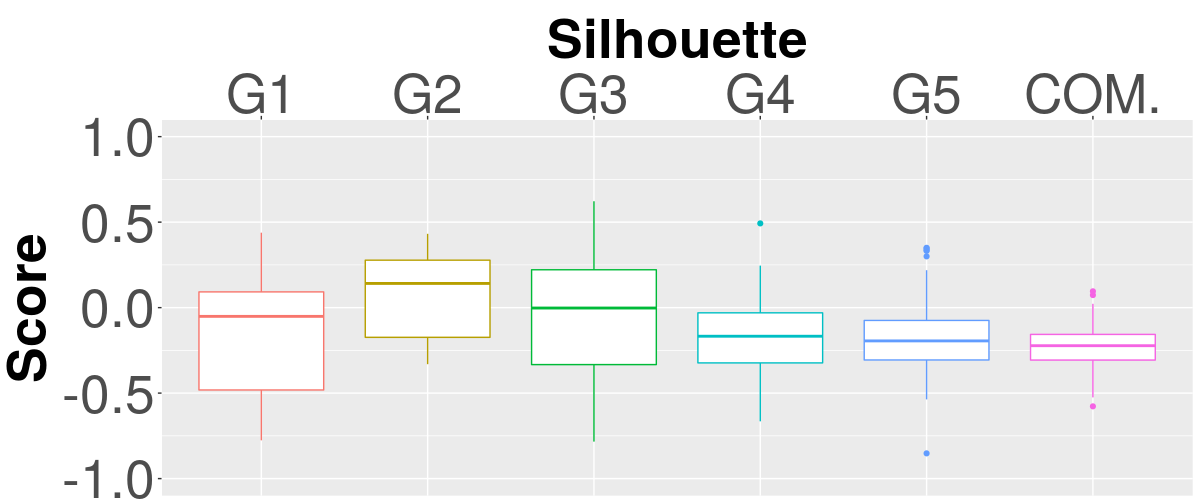}} \\
			
		\end{tabular} 
		\caption{Compactness and silhouette scores.} 
		\label{fig:comparison}
	\end{figure*}

	
	\rev{We compute and report for each set the corresponding compactness and silhouette scores.} Figure~\ref{fig:CompactnessDAG}, Figure~\ref{fig:CompactnessDP1AP1GP1}, and Figure~\ref{fig:CompactnessG1G2G3G4G5} report the compactness scores computed for the three question sets.

	As it can be seen in Figure~\ref{fig:CompactnessDAG}, using \textbf{A} as input yields the most compact clusters. 
	In particular, most of the scores are smaller than 40. When domain specific questions (\textbf{D}) are used as the features, we also obtain low compactness scores, albeit being larger than using \textbf{A}. If only generic questions, \ie \textbf{G}, are utilized, worse clustering solutions are seen. When comparing the results obtained by using \textbf{G} with those of using \textbf{G+AP2}, we can see that adding \textbf{AP2} to \textbf{G} contributes to a better clustering. 
	%
	%
	Concerning QS$_2$ where only privacy relevant questions are considered, we see that using domain specific privacy relevant questions (\textbf{DP1}) allows us to gain the most discriminative clusters. \rev{Using the subset of privacy relevant questions, \ie \textbf{AP1} is also beneficial to the clustering of user profiles.}
	
	%
	For QS$_3$, there are comparable clustering solutions when using the features sets \textbf{G$_1$}, \textbf{G$_3$}, \textbf{G$_4$}, and \textbf{G$_5$}. The best clustering is obtained with \textbf{G$_2$}.
	%
	
	%



	The silhouette scores in Figure~\ref{fig:SilhouetteDAG}, Figure~\ref{fig:SilhouetteDP1AP1GP1}, and Figure~\ref{fig:SilhouetteG1G2G3G4G5} further enforce the compactness ones. \textbf{A} is the feature set that achieves the best silhouette for QS$_1$. Adding \textbf{AP2} to \textbf{G} helps achieve a better clustering solution, compared to using only \textbf{G}.
	
	\begin{tcolorbox}[boxrule=0.86pt,left=0.3em, right=0.3em,top=0.1em, bottom=0.05em]
		\textbf{Answer to RQ$_{2}$.} According to the performed evaluation, generic questions plus generalizable ones (i.e., \textbf{G+AP2}) provide the best clustering solution.
	\end{tcolorbox}

	\subsection{\rqthird} 
	
	An issue with clustering is whenever there is a new user to be classified, it is necessary to re-run the whole process. \rev{This is a time consuming phase, especially where there is a large number of users. Thus, we propose a more feasible way to assign new users to clusters, avoiding repetitive clustering.} 
	
	\begin{figure}[h!]
		\centering
		\includegraphics[width=0.60\textwidth]{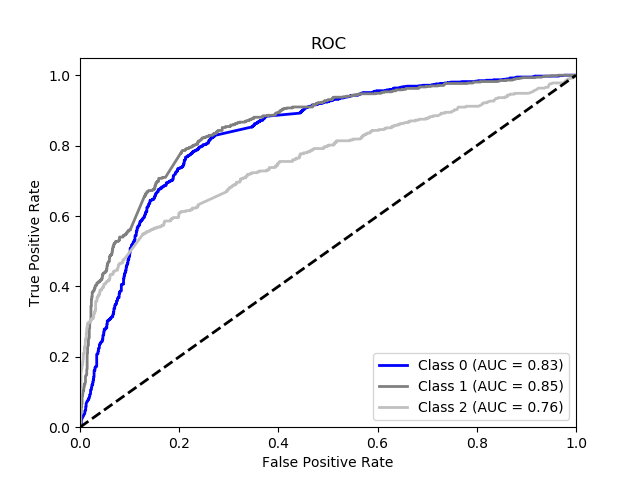}
		\caption{ROC curves, three categories.}
		\label{fig:ROC3}
	\end{figure}

	\rev{Given that there is an existing categorization of user profiles, the feed-forward neural network presented in Section~\ref{sec:NeuralNetwork} is used to classify a new user into a suitable group. 
		Once clusters have been obtained, we feed them as input to train the neural network and perform the testing using the ten-fold cross-validation procedure. It is worth mentioning that we use three clusters instead of four as explained in Section~\ref{sec:Categorization}.} 
	
	\rev{The final performance measured by means of ROC curves is depicted in Figure~\ref{fig:ROC3}. In particular, the AUC values for Class $0$, Class $1$, and Class $2$ are $0.83$, $0.85$, and $0.76$, respectively. The curves representing the three classes reside near the upper left corner, implying a good prediction performance. Overall, the curves and the AUC values demonstrate that the obtained performance is much better compared to that before clustering in Figure~\ref{fig:ROC1} and Figure~\ref{fig:ROC2}.} This suggests that properly clustering user profiles can substantially increase the neural network's prediction performance.

	

	Next, we validate the performance of \PR as follows. We opted for the ten-fold cross validation technique~\citep{Kohavi:1995:SCB:1643031.1643047}, where the dataset is split into ten equal folds, and the evaluation is done in ten rounds. \rev{By each round, one fold is utilized as testing, and the other nine folds are merged to create the training data.} 
	In a testing fold, for each user, the features are split into two 
	parts, one part is fed as query, and the remaining part is removed to be used as ground-truth data. %
	The ratio of the number of settings used as query to the total number of settings is called $\alpha$. This simulates a real scenario, where the user already specified some settings, and the system is expected to \rev{recommend the rest, corresponding to 
		the ground-truth data. For each user, \PR returns a ranked list of $N$ settings ($N$ is configurable), and the evaluation metrics are computed on the test set as follows. The recommended items are then compared with the ground-truth data to evaluate the performance. Eventually, we average out the metrics obtained from the testing folds to produce the final results.}
	
	We experiment with different configurations by varying $\alpha$, $k$: the number of neighbor users used for the computation, $N$: the number of recommended items. In particular, $\alpha= \{0.1,0.3,0.5\}$; $k = \{3,5,10,15\}$; and $N$ is varied from 1 to 50, simulating a real-world scenario where 
	users have to set several settings. 
	The precision-recall curves are then sketched 
	following these parameters.
	
	
	\begin{figure}[h!]
		\centering
		\includegraphics[width=0.68\textwidth]{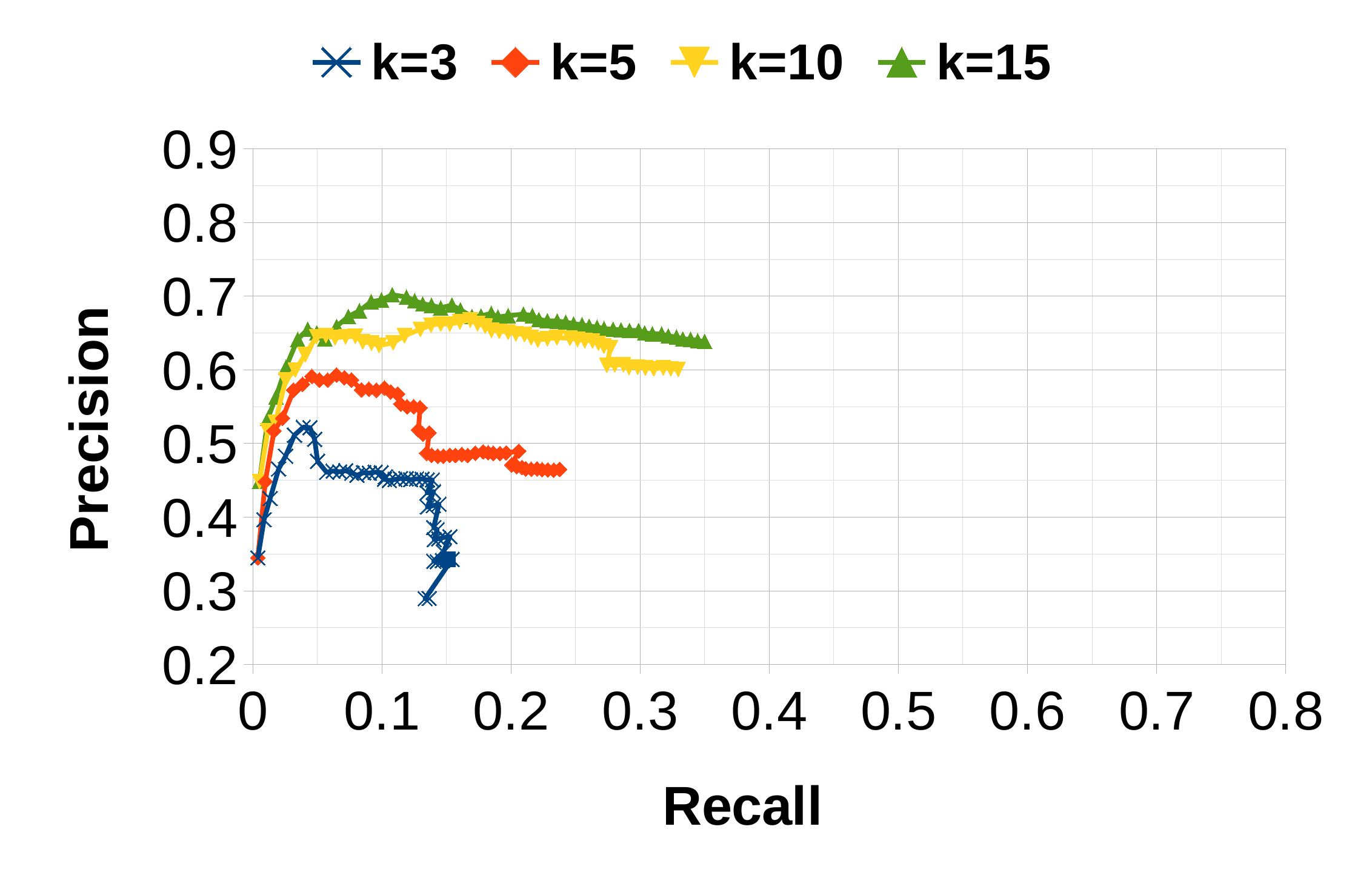}
		\caption{Configuration C$_1$.}
		\label{fig:PrecisionRecall_C1}
	\end{figure}

	As seen in Figure~\ref{fig:PrecisionRecall_C1}, when $\alpha=0.1$, \ie only a small amount of data is 
	used as query, \PR recommends relevant settings to users, however with considerably low precision and recall. \rev{For instance, when $k=3$, a maximum precision of 0.52 is obtained 
		and the maximum precision is $0.7$ when $k=15$. Similarly, the recall scores are low, \ie smaller than 0.4 by all the configurations. Altogether, this implies a mediocre performance which is understandable as the configuration with $\alpha=0.1$ corresponds to the case where the user only specified a few settings, and the system has limited context to recommend additional settings.} 
	
	
	%

	
	\begin{figure}[h!]
		\centering
		\includegraphics[width=0.68\textwidth]{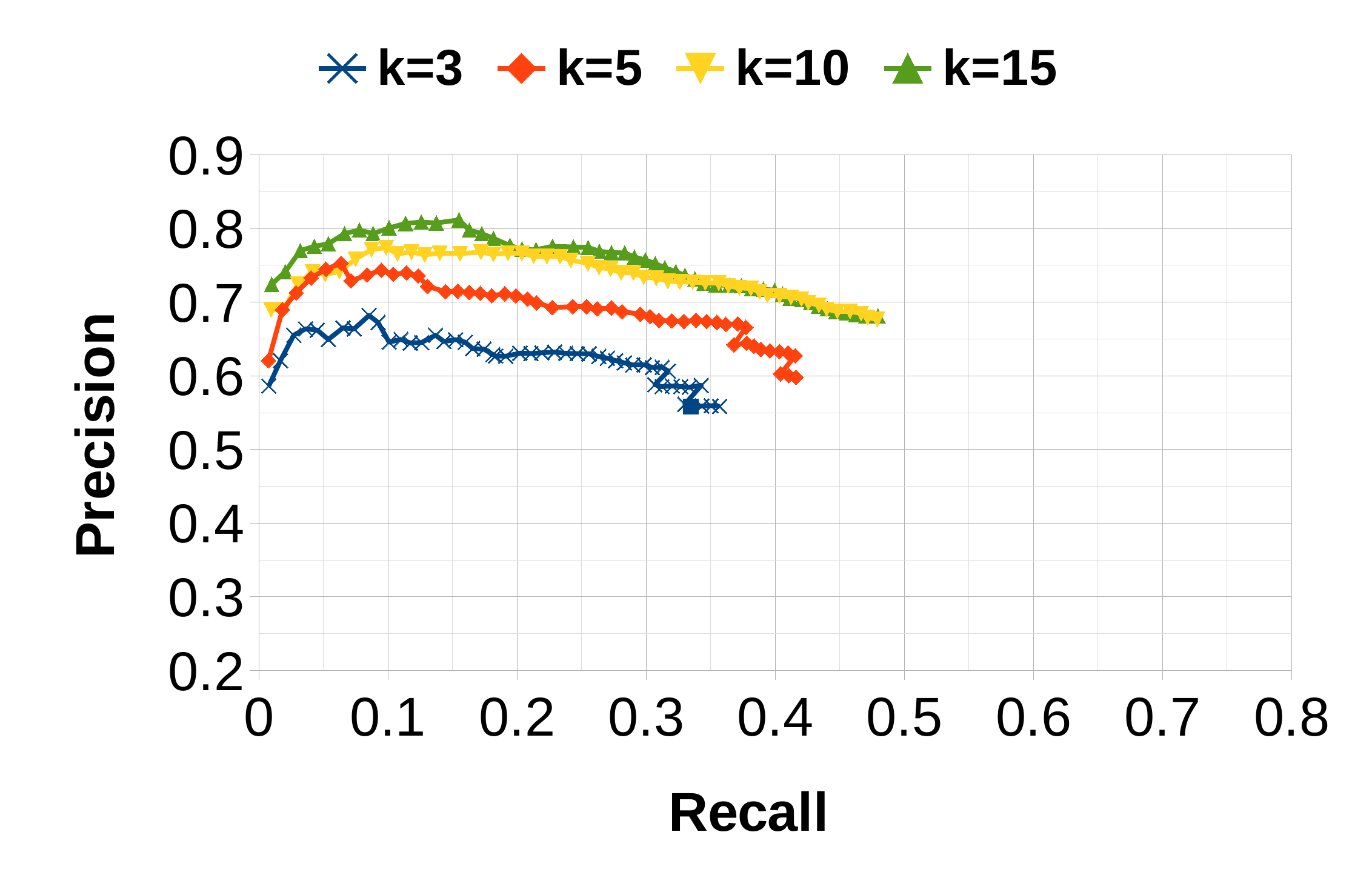}
		\caption{Configuration C$_2$.}
		\label{fig:PrecisionRecall_C2}
	\end{figure}

	When we increase $\alpha$ to $0.3$, there is an improvement in both precision and recall as in Figure~\ref{fig:PrecisionRecall_C2}, \rev{compared to the results obtained with $\alpha=0.1$ in Figure~\ref{fig:PrecisionRecall_C1}}. 
	Precision scores are always larger than 0.55 in the configurations, with 0.80 being the maximum value. \rev{Similarly, we also see that recall scores are gradually improved. For instance, a maximum recall of 0.35 is achieved with $k=3$, and the corresponding maximum for $k=15$ is $0.48$.}

	\begin{figure}[h!]
		\centering
		\includegraphics[width=0.68\textwidth]{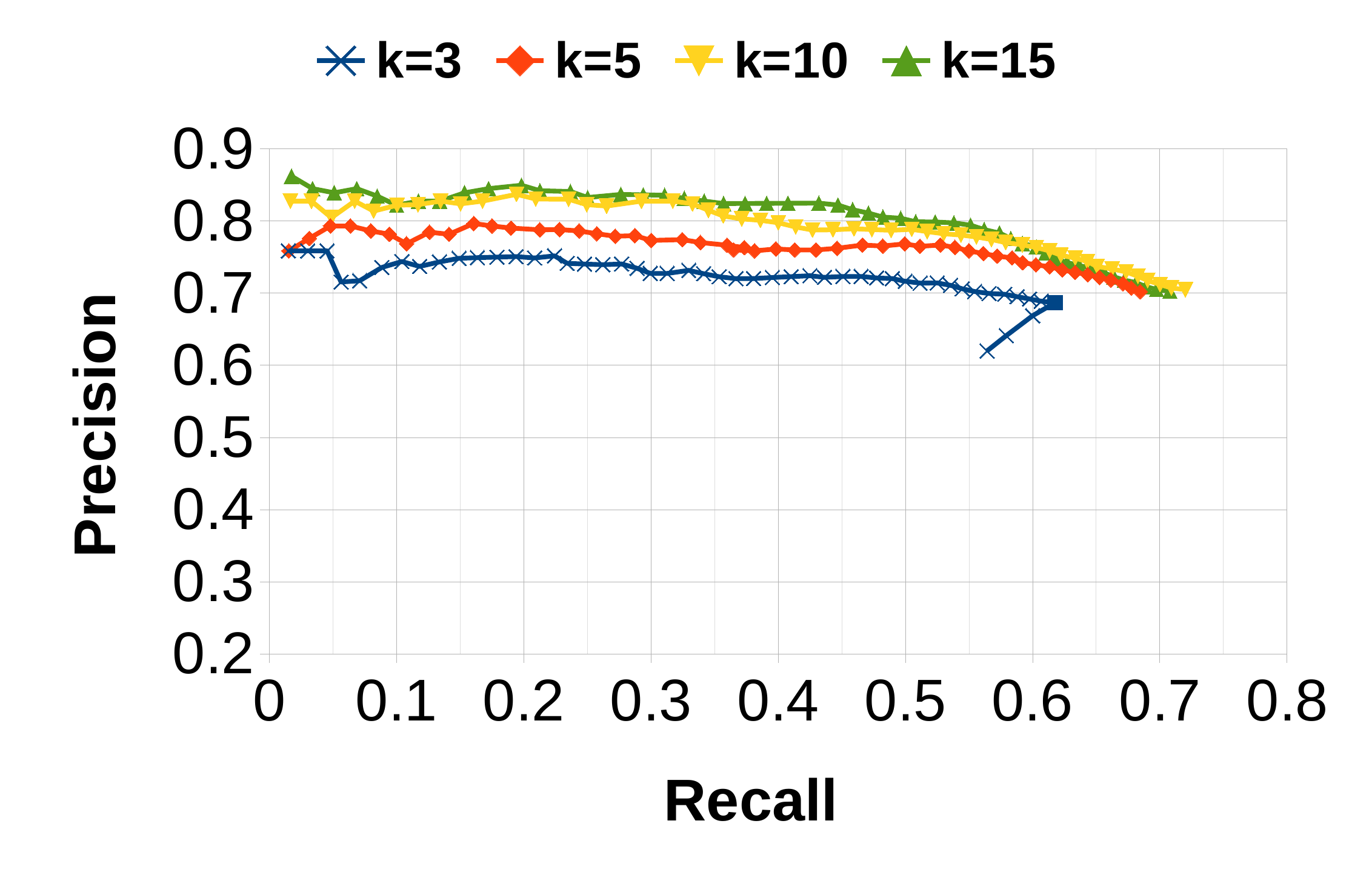}
		\caption{Configuration C$_3$.}
		\label{fig:PrecisionRecall_C3}
	\end{figure}
	
	Such an improvement is more evident when $\alpha=0.50$, \rev{\ie a half of the settings is used as query}. In Figure~\ref{fig:PrecisionRecall_C3}, apart from some outliers, most of the precision scores are larger than 0.70, with 0.85 as the maximum value. Compared to the previous configurations with $\alpha=0.1$ and $\alpha=0.3$, the recall scores are also better, \ie with a longer list of items, recall increases substantially. \rev{In particular, a recall of $0.73$ is seen when $k=10$ and $k=15$.}

	Concerning the number of neighbors used for computing recommendations, \rev{\ie $k$ (see Section~\ref{sec:RecSys} and Formula \ref{eqn:Prediction})}, by considering Figure~\ref{fig:PrecisionRecall_C1}, Figure~\ref{fig:PrecisionRecall_C2}, and Figure~\ref{fig:PrecisionRecall_C3} together, it is evident that adding more users for the computation contributes to a better prediction performance. For instance, by increasing k from $3$ to $5$, $10$, and $15$, we boost both precision and recall by all the cut-off values $N$.

	Altogether, the experimental results show 
	that even if users perceive their categories differently as shown in \textbf{RQ$_1$}, once we have identified their right privacy group  
	\PR can exploit the categories to provide relevant settings to users, \rev{though the considered dataset is pretty small. We anticipate that its performance can be further enhanced, if there is more data for training.}
	
	\begin{tcolorbox}[boxrule=0.86pt,left=0.3em, right=0.3em,top=0.1em, bottom=0.05em]
		\textbf{Answer to RQ$_3$.} \PR recommends highly relevant settings to a user, though there is limited amount of data available for training. The prediction performance improves alongside the amount of data fed as input.
	\end{tcolorbox}

	\section{Discussion}	
	\label{sec:Discussion}

	\rev{This section provides discussion related to the possible extensions of our work, as well as the threats to validity of our findings.}

	\subsection{\rev{Extendability}}
	
	\vspace{.1cm}
	\noindent \rev{\textbf{Dataset.} In our work, we utilized a small dataset for the evaluation. The amount of training data may impact the performance of both the clustering and classification phases. Moreover, as \PR is a collaborative-filtering recommender system, its performance is heavily driven by the quality and amount of data. We anticipate that we may need to calibrate the systems' parameters to maintain both timing efficiency and effectiveness with more data.} 
	
	\vspace{.1cm}
	\noindent \rev{\textbf{The unsupervised algorithm.} In the scope of this paper, we used the K-Medoids algorithm to cluster the user profiles. Such a technique has been chosen due to its simplicity and effectiveness. In fact, several clustering algorithms could be employed to categorize user profiles. Thus, the outcome of a clustering solution depends heavily on the considered techniques. We plan to extend our work by considering other clustering algorithms, such as CLARA~\citep{books/wi/KaufmanR90}, or DBSCAN~\citep{10.5555/3001460.3001507}.}

	\vspace{.1cm}
	\noindent \rev{\textbf{The supervised classifier.} The neural network used to classify user profiles may be suitable only for the considered dataset. For a different dataset, it is necessary to find adequate network configurations employing an empirical evaluation. For instance, the number of hidden layers, or the number of neurons for each layer, should be considerably increased to deal with a larger number of user profiles.}

	\subsection{Threats to validity}
	
	\rev{We are aware of the existence of some threats that might harm the validity of the performed experiments as they are presented as follows.} 
	
	\begin{itemize}
		\item Threats to \emph{construct validity} are related to any factor that can compromise the validity of the given observations. The main threat to construct validity is related to the size of the analyzed data. The used dataset is indeed relatively small but has the advantage of coming from a recent work \citep{sanchez_recommendation_2020} thus reflecting users' contemporary privacy behaviors. More extensive experiments are under planning encompassing other ethical dimensions beyond privacy.
		\item Concerning the threats to \emph{internal validity}, \ie any confounding factor that could influence our findings, we attempted to avoid any bias in the automatic creation of user profiles and in the way we split the full data into groups. We tried to mitigate this threat by semantically analyzing and double-checking the clusters obtained by the proposed approach.
		\item Concerning the threats to \emph{external validity}, they are related to the generalizability of our results. This is about checking the adequacy of our privacy profiles in other contexts, notably, in the traveling or IoT domains. Generalizability is actually our initial driver for extracting privacy profiles from general moral questions. Thus, further experimental evidence is planned to support the reported paper results.
	\end{itemize}

	\section{Related work}
	\label{sec:RelatedWork}

	\rev{This section reviews the related work and their main characteristics to position our approach in the current scenario for eliciting, profiling, and predicting user privacy preferences.}
	
	\rev{\subsection{Overview}}
	
	The work presented in this paper has been done in the context of the design of the EXOSOUL research project that aims at providing users with a personalized software layer that mediates users' interactions with the digital world according to user's ethics, including privacy preferences~\citep{anonymized_project,AutiliRIPT19}. 
	
	\rev{According to various studies~\citep{milne2004strategies,obar2020biggest}, the vast majority of users do not bother to read privacy agreements because of the excessive language and confusing explanations~\citep{bhatia2016theory,jensen2004privacy,mcdonald2009comparative,reidenberg2015disagreeable}; it is unreasonable also to expect they will read them on a regularly basis~\citep{mcdonald2008cost}. Resignation from privacy choices may also be a result of their dissatisfaction with the lack of options and excessive complexity~\citep{colnago2020informing}.}
	
	\rev{Privacy profiling is at the core of our work therefore, most related studies are on user clustering, privacy profiling, and privacy preferences settings.} The more significant part of existing studies about privacy profiling develop on the work of Westin~\citep{westin_biblio}. Based on a series of privacy-related surveys, the author established ``\emph{Privacy Indexes}'' for most of these polls to summarize results, indicate trends in privacy concerns, and suggest a widely recognized segmentation methodology of ``\emph{Privacy Profiles.}''\rev{The methodology he applied classifies people into three categories: \textit{privacy fundamentalists, pragmatists, and unconcerned}. Because of the commercial nature of Westin's surveys, the methodology and the details of how privacy indexes were calculated are not fully disclosed, so we rely on subsequent works \citep{Kumaraguru_Cranor:2005} that deeply analyzed and reported them.}
	
	
	\rev{Westin Segmentation, and particularly the pragmatism adherence of the consumers were criticized by the work of Hoofnagle and Urban~\citep{hoofnagle2014alan,urban2014privacy}. Their experimental work investigated customer expectations for privacy safeguards, showing that many people believe they have greater protection than they really do due to a lack of knowledge about corporate practices, privacy policies, and data usage limits.} Westin's methodology has been applied\rev{, revised, and expanded }in several empirical studies on privacy that include collected data. The categorizations that are most relevant to our research are reported in Table~\ref{tab:Categories}.
	\rev{
		Dupree \etal \citep{10.1145/2858036.2858214} analyzed data from surveys and participants interviews, the authors identified five user clusters that emerge from end-user behaviors, including \emph{Fundamentalists}, \emph{Lazy Experts}, \emph{Technicians}, \emph{Amateurs} and the \emph{Marginally Concerned}.
		Schairer \etal \citep{10.1093/jamia/ocz010} presented a model of privacy disposition and its development based on qualitative research on privacy considerations in the context of emerging health technologies. The authors identified six clusters, including \emph{Fatalism}, \emph{Nothing to hide}, \emph{Something to hide}, \emph{Tradeoff}, \emph{Personal responsibility}, \emph{Moral right}.
		In the research proposed by Sanchez \etal \citep{sanchez_recommendation_2020} the authors presented the results of a fitness-related simulation and questionnaire 
		to classify users according to their privacy-related preferences. They used two different sets of labels for their clusters, one for the computed privacy-profile assignment consisting of six groups and one for self-assessment they proposed to the users consisting in four groups. The first clusters were labeled as: \emph{Unconcerned}, \emph{Socially Active}, \emph{Health-focused}, \emph{Minimal}, \emph{Anonymous}, \emph{Strict}. The second clusters were labeled as: \emph{Privacy Conservative}, \emph{Unconcerned}, \emph{Fence-Sitter}, \emph{Advanced User}. 
	}
	
	\rev{ \subsection{Profiling and Clusterization} }
	
	\rev{Different approaches were used in recent works to create user profiling starting from data collection and analysis. 
		Lee and Kobsa~\citep{7845392} performed a cluster analysis on online survey data composed of IoT scenarios and user responses like reaction parameters. Because all parameters have either categorical or ordinal values, the authors utilized K-modes, a variant of the K-means clustering algorithm.
		Qin \etal~\citep{10.1007/978-3-540-68825-9_22} proposed a user's preferences prediction through clusterization of partial preference relations on the MovieLens dataset, which is commonly used to test collaborative filtering technology.
		According to Fernquist \etal~\citep{fernquist2017iot}, users may also be identified by their data profiles created by their device based on time and events: researchers gathered information on how and when people use their networked devices, recording the time period in which a user interacts or transmits data and the specific place. For the sake of interpreting their findings, the researchers took into account three distinct sorts of events: voice calls, texts, and data transfers, as well as combinations of these. The findings showed that the profiles studied may be used to identify the user.}
	
	
	
	\rev{\subsection{Automating privacy settings} }
	
	Concerning automating privacy preferences settings, the closest approach is by Liu \etal~\citep{lin2014modeling,Liu2020_PHD} and by the Personalized Privacy Assistant Project team~\citep{sadeh2021personalized}. Their approach employs user categorizations that are obtained by mining existing privacy settings in the app domain, complemented with an initial dialogue with the user to select the appropriate profile. Our approach is also based on privacy profiles. However, they are obtained by analyzing data resulting from questions that relate to the user's ethics and are not concerned with any specific domain. 
	
	\rev{Wilson \etal~\citep{wilson2013privacy} identified the impact of privacy profiles on the preferences, sharing inclinations, and overall satisfaction levels of users of location-sharing apps. Their findings demonstrate that privacy profiles for location sharing settings can have a long-lasting impact on how users perceive their privacy, even in the face of ongoing opportunities to reflect on the sharing outcomes that result from their chosen settings. This implies that attempts to simplify privacy settings should be performed with caution since such simplicity may easily impact the people with whom the settings are intended to interact and educate.}
	
	Brandimarte~\etal~\citep{brandimarte2013misplaced} investigated the concept that giving people a greater sense of control over the release and access to private information -- even information that enables them to be personally identified -- would improve their willingness to provide sensitive information. If their desire to reveal adequately rises, this control may, counter-intuitively, result in being more slack. 
	
	In their research, participants in a publication were informed that a profile comprising their information would be produced automatically and published online once the website was finished. Other participants were informed that only half of their profiles would be published online. The uncertain publishing condition was designed to reduce participants' sense of control over the public distribution of their survey responses without actually decreasing access by others. Their theories predicted that decreasing control would limit the desire to reveal in the uncertain publishing condition, notwithstanding lower external costs or hazards. According to the researchers' results, if individuals behave in enough offsetting manner, devices supposed to safeguard them might instead end up worsening the hazards they confront.

	\rev{\subsection{Surveys and regulations}}
	
	\rev{Based on previous research in survey technique and related domains, Redmiles \etal~\citep{redmiles2017summary} provide a set of important recommendations for conducting self-report usability studies. There are established criteria and suggestions for collecting good quality self-report data in other sectors that depend on self-report data, such as health and social sciences. We used this information as a guideline for selecting and refining question groups.}
	
	\rev{As discussed by Emami-Naeini~\etal~\citep{emami2020ask}, surveys and interviews can be administered with consolidated methodologies like the \textit{Delphi Method}. This method is ``\emph{a method for the systematic solicitation and collection of judgments on a particular topic through a set of carefully designed sequential questionnaires interspersed with summarized information and feedback of opinions derived from earlier responses}''~\citep{atherton1976group}. Using a three-round Delphi process, the authors conducted an expert elicitation study with 22 privacy and security experts to identify the factors that experts believe are important for consumers to consider when comparing the privacy and security of IoT devices to inform their purchase decisions. The same methodology could be used to elicit preferences from the users.}
	
	\rev{Considering the research theme, we took into account the General Data Protection Regulation (GDPR) \citep{GDPR2016regulation}, the document that governs the storing, processing, and use of personal data by the European Union (EU) as of May 25, 2018. Even if not based in the EU, the GDPR applies to all third parties that operate in the EU market or access the data of EU citizens.}

	\vspace{.2cm}

	\section{Conclusion and future work}
	\label{sec:Conclusions}

	
	This paper proposes a holistic approach consisting of both supervised and unsupervised learning to identify privacy profiles. By finding a set of questions suitable for assessing privacy profiles, we recommend relevant privacy preferences to users.
	An empirical study on the proposed system using a fitness dataset shows that generic questions are suitable for categorizing user profiles and recommending privacy settings. For future work, besides developing further experimental evidence supporting the results reported in this paper, we will work in the direction of building user profiles that cover other ethical dimensions beyond privacy. Last but not least, in the scope of the Exosoul project, we will deploy the conceived techniques to analyze data collected from users, studying the characteristics of users' behaviors and their attitudes in the digital world.

	

	\section*{Acknowledgements}
	The research described in this paper has been carried out as part of the EXOSOUL project and 
	also partially supported by the Centre of excellence EX-EMERGE (Centre of EXcellence on Connected, Geo-localized and Cyber-secure vehicles) of the University of L'Aquila, Italy.

	
	\bibliographystyle{elsarticle-num}
	\bibliography{main}

\begin{thebibliography}{10}
\expandafter\ifx\csname url\endcsname\relax
  \def\url#1{\texttt{#1}}\fi
\expandafter\ifx\csname urlprefix\endcsname\relax\def\urlprefix{URL }\fi
\expandafter\ifx\csname href\endcsname\relax
  \def\href#1#2{#2} \def\path#1{#1}\fi

\bibitem{westin_biblio}
A.~F. Westin,
  \href{https://web.archive.org/web/20051224000944/http://www.privacyexchange.org/iss/surveys/surveybibliography603.pdf}{Bibliography
  of surveys of the u.s. public, 1970-2003} (2003).
\newline\urlprefix\url{https://web.archive.org/web/20051224000944/http://www.privacyexchange.org/iss/surveys/surveybibliography603.pdf}

\bibitem{Kumaraguru_Cranor:2005}
P.~Kumaraguru, L.~F. Cranor, {P}rivacy {I}ndexes: {A} {S}urvey of {W}estin's
  {S}tudies, Tech. Rep. CMU-ISRI-5-138, Institute for Software Research
  International, School of Computer Science, Carnegie Mellon University,
  Pittsburgh, PA (Dezember 2005).

\bibitem{ZHAO2019449}
Y.~Zhao, Y.~Yu, Y.~Li, G.~Han, X.~Du,
  \href{https://www.sciencedirect.com/science/article/pii/S0020025518309174}{Machine
  learning based privacy-preserving fair data trading in big data market},
  Information Sciences 478 (2019) 449--460.
\newblock \href {https://doi.org/https://doi.org/10.1016/j.ins.2018.11.028}
  {\path{doi:https://doi.org/10.1016/j.ins.2018.11.028}}.
\newline\urlprefix\url{https://www.sciencedirect.com/science/article/pii/S0020025518309174}

\bibitem{10.1007/978-3-540-68825-9_22}
M.~Qin, S.~Buffett, M.~W. Fleming, Predicting user preferences via
  similarity-based clustering, in: S.~Bergler (Ed.), Advances in Artificial
  Intelligence, Springer Berlin Heidelberg, Berlin, Heidelberg, 2008, pp.
  222--233.

\bibitem{7845392}
H.~Lee, A.~Kobsa, Understanding user privacy in internet of things
  environments, in: 2016 IEEE 3rd World Forum on Internet of Things (WF-IoT),
  2016, pp. 407--412.
\newblock \href {https://doi.org/10.1109/WF-IoT.2016.7845392}
  {\path{doi:10.1109/WF-IoT.2016.7845392}}.

\bibitem{lin2014modeling}
J.~Lin, B.~Liu, N.~Sadeh, J.~I. Hong, Modeling users’ mobile app privacy
  preferences: Restoring usability in a sea of permission settings, in: 10th
  Symposium On Usable Privacy and Security ($\{$SOUPS$\}$ 2014), 2014, pp.
  199--212.

\bibitem{sanchez_recommendation_2020}
O.~R. Sanchez, I.~Torre, Y.~He, B.~P. Knijnenburg,
  \href{https://doi.org/10.1007/s11257-019-09246-3}{A recommendation approach
  for user privacy preferences in the fitness domain}, User Modeling and
  User-Adapted Interaction 30~(3) (2020) 513--565.
\newblock \href {https://doi.org/10.1007/s11257-019-09246-3}
  {\path{doi:10.1007/s11257-019-09246-3}}.
\newline\urlprefix\url{https://doi.org/10.1007/s11257-019-09246-3}

\bibitem{Liu2020_PHD}
B.~Liu,
  \href{https://kilthub.cmu.edu/articles/thesis/Can_Machine_Learning_Help_People_Configure_Their_Mobile_App_Privacy_Settings_/11591340}{{Can
  Machine Learning Help People Configure Their Mobile App Privacy Settings?}}
  (1 2020).
\newblock \href {https://doi.org/10.1184/R1/11591340.v1}
  {\path{doi:10.1184/R1/11591340.v1}}.
\newline\urlprefix\url{https://kilthub.cmu.edu/articles/thesis/Can_Machine_Learning_Help_People_Configure_Their_Mobile_App_Privacy_Settings_/11591340}

\bibitem{McKinsey2020}
McKinsey, \href{https://mck.co/3trP4OV}{How COVID-19 has pushed companies over
  the technology tipping point—and transformed business forever
  (https://mck.co/3trP4OV)}, 2020.
\newline\urlprefix\url{https://mck.co/3trP4OV}

\bibitem{/content/publication/bb167041-en}
OECD, \href{https://www.oecd-ilibrary.org/content/publication/bb167041-en}{OECD
  Digital Economy Outlook 2020}, 2020.
\newblock \href
  {https://doi.org/https://doi.org/https://doi.org/10.1787/bb167041-en}
  {\path{doi:https://doi.org/https://doi.org/10.1787/bb167041-en}}.
\newline\urlprefix\url{https://www.oecd-ilibrary.org/content/publication/bb167041-en}

\bibitem{PUJAHARI2019126}
A.~Pujahari, D.~S. Sisodia,
  \href{https://www.sciencedirect.com/science/article/pii/S0020025519302774}{Modeling
  side information in preference relation based restricted boltzmann machine
  for recommender systems}, Information Sciences 490 (2019) 126--145.
\newblock \href {https://doi.org/https://doi.org/10.1016/j.ins.2019.03.064}
  {\path{doi:https://doi.org/10.1016/j.ins.2019.03.064}}.
\newline\urlprefix\url{https://www.sciencedirect.com/science/article/pii/S0020025519302774}

\bibitem{liu_follow_2016}
B.~Liu, M.~S. Andersen, F.~Schaub, H.~Almuhimedi, S.~Zhang, N.~Sadeh,
  A.~Acquisti, Y.~Agarwal, Follow my recommendations: A personalized privacy
  assistant for mobile app permissions, in: Proceedings of the Twelfth {USENIX}
  Conference on Usable Privacy and Security, {SOUPS} '16, {USENIX} Association,
  2016, pp. 27--41, event-place: Denver, {CO}, {USA}.

\bibitem{AutiliRIPT19}
M.~Autili, D.~D. Ruscio, P.~Inverardi, P.~Pelliccione, M.~Tivoli,
  \href{https://doi.org/10.1109/ACCESS.2019.2916203}{A software exoskeleton to
  protect and support citizen's ethics and privacy in the digital world},
  {IEEE} Access 7 (2019) 62011--62021.
\newblock \href {https://doi.org/10.1109/ACCESS.2019.2916203}
  {\path{doi:10.1109/ACCESS.2019.2916203}}.
\newline\urlprefix\url{https://doi.org/10.1109/ACCESS.2019.2916203}

\bibitem{10.1145/2858036.2858214}
J.~L. Dupree, R.~Devries, D.~M. Berry, E.~Lank,
  \href{https://doi.org/10.1145/2858036.2858214}{Privacy personas: Clustering
  users via attitudes and behaviors toward security practices}, in: Proceedings
  of the 2016 CHI Conference on Human Factors in Computing Systems, CHI '16,
  Association for Computing Machinery, New York, NY, USA, 2016, p. 5228–5239.
\newblock \href {https://doi.org/10.1145/2858036.2858214}
  {\path{doi:10.1145/2858036.2858214}}.
\newline\urlprefix\url{https://doi.org/10.1145/2858036.2858214}

\bibitem{10.1093/jamia/ocz010}
C.~E. Schairer, C.~Cheung, C.~Kseniya~Rubanovich, M.~Cho, L.~F. Cranor, C.~S.
  Bloss, \href{https://doi.org/10.1093/jamia/ocz010}{{Disposition toward
  privacy and information disclosure in the context of emerging health
  technologies}}, Journal of the American Medical Informatics Association
  26~(7) (2019) 610--619.
\newblock \href
  {http://arxiv.org/abs/https://academic.oup.com/jamia/article-pdf/26/7/610/34151415/ocz010.pdf}
  {\path{arXiv:https://academic.oup.com/jamia/article-pdf/26/7/610/34151415/ocz010.pdf}},
  \href {https://doi.org/10.1093/jamia/ocz010}
  {\path{doi:10.1093/jamia/ocz010}}.
\newline\urlprefix\url{https://doi.org/10.1093/jamia/ocz010}

\bibitem{PARK20093336}
H.-S. Park, C.-H. Jun,
  \href{https://www.sciencedirect.com/science/article/pii/S095741740800081X}{A
  simple and fast algorithm for k-medoids clustering}, Expert Systems with
  Applications 36~(2, Part 2) (2009) 3336--3341.
\newblock \href {https://doi.org/https://doi.org/10.1016/j.eswa.2008.01.039}
  {\path{doi:https://doi.org/10.1016/j.eswa.2008.01.039}}.
\newline\urlprefix\url{https://www.sciencedirect.com/science/article/pii/S095741740800081X}

\bibitem{10.5555/523781}
K.~Gurney, An Introduction to Neural Networks, Taylor and Francis, Inc., USA,
  1997.

\bibitem{Bishop:1995:NNP:525960}
C.~M. Bishop, Neural networks for pattern recognition, Oxford University Press,
  Inc., New York, NY, USA, 1995.

\bibitem{RePEc:eee:intfor:v:14:y:1998:i:1:p:35-62}
G.~Zhang, B.~Eddy~Patuwo, M.~Y.~Hu, Forecasting with artificial neural
  networks:: {The} state of the art, International Journal of Forecasting
  14~(1) (1998) 35--62.

\bibitem{Schafer:2007:CFR:1768197.1768208}
J.~B. Schafer, D.~Frankowski, J.~Herlocker, S.~Sen, The adaptive web,
  Springer-Verlag, Berlin, Heidelberg, 2007, pp. 291--324.

\bibitem{Aggarwal2016}
C.~Aggarwal, Neighborhood-based collaborative filtering, in: Recommender
  systems: {The} textbook, Springer International Publishing, Cham, 2016, pp.
  29--70.

\bibitem{DBLP:conf/rweb/NoiaO15}
T.~D. Noia, V.~C. Ostuni,
  \href{https://doi.org/10.1007/978-3-319-21768-0\_4}{Recommender systems and
  linked open data}, in: W.~Faber, A.~Paschke (Eds.), Reasoning Web. Web Logic
  Rules - 11th International Summer School 2015, Berlin, Germany, July 31 -
  August 4, 2015, Tutorial Lectures, Vol. 9203 of Lecture Notes in Computer
  Science, Springer, 2015, pp. 88--113.
\newblock \href {https://doi.org/10.1007/978-3-319-21768-0\_4}
  {\path{doi:10.1007/978-3-319-21768-0\_4}}.
\newline\urlprefix\url{https://doi.org/10.1007/978-3-319-21768-0\_4}

\bibitem{sadeh2021personalized}
N.~Sadeh, B.~Liu, A.~Das, M.~Degeling, F.~Schaub, Personalized privacy
  assistant, uS Patent 10,956,586 (Mar.~23 2021).

\bibitem{10.1109/ICDM.2010.35}
Y.~Liu, Z.~Li, H.~Xiong, X.~Gao, J.~Wu,
  \href{https://doi.org/10.1109/ICDM.2010.35}{Understanding of internal
  clustering validation measures}, in: Proceedings of the 2010 IEEE
  International Conference on Data Mining, ICDM '10, IEEE Computer Society,
  USA, 2010, p. 911–916.
\newblock \href {https://doi.org/10.1109/ICDM.2010.35}
  {\path{doi:10.1109/ICDM.2010.35}}.
\newline\urlprefix\url{https://doi.org/10.1109/ICDM.2010.35}

\bibitem{10.1016/j.patrec.2005.10.010}
T.~Fawcett, \href{https://doi.org/10.1016/j.patrec.2005.10.010}{An introduction
  to roc analysis}, Pattern Recogn. Lett. 27~(8) (2006) 861--874.
\newblock \href {https://doi.org/10.1016/j.patrec.2005.10.010}
  {\path{doi:10.1016/j.patrec.2005.10.010}}.
\newline\urlprefix\url{https://doi.org/10.1016/j.patrec.2005.10.010}

\bibitem{10.1145/3439726}
S.~Minaee, N.~Kalchbrenner, E.~Cambria, N.~Nikzad, M.~Chenaghlu, J.~Gao,
  \href{https://doi.org/10.1145/3439726}{Deep learning--based text
  classification: A comprehensive review}, ACM Comput. Surv. 54~(3) (Apr.
  2021).
\newblock \href {https://doi.org/10.1145/3439726} {\path{doi:10.1145/3439726}}.
\newline\urlprefix\url{https://doi.org/10.1145/3439726}

\bibitem{1311506}
J.~A. Burns, G.~M. Whitesides, Feed-forward neural networks in chemistry:
  Mathematical systems for classification and pattern recognition, Chem. Rev.
  93 (1993) 2583--2601, 377.

\bibitem{8906979}
P.~T. {Nguyen}, J.~{Di Rocco}, D.~{Di Ruscio}, A.~{Pierantonio}, L.~{Iovino},
  Automated classification of metamodel repositories: A machine learning
  approach, in: 2019 ACM/IEEE 22nd International Conference on Model Driven
  Engineering Languages and Systems (MODELS), 2019, pp. 272--282.
\newblock \href {https://doi.org/10.1109/MODELS.2019.00011}
  {\path{doi:10.1109/MODELS.2019.00011}}.

\bibitem{Kohavi:1995:SCB:1643031.1643047}
R.~Kohavi, A study of cross-validation and bootstrap for accuracy estimation
  and model selection, in: Proceedings of the 14th international joint
  conference on artificial intelligence - volume 2, {IJCAI}'95, Morgan Kaufmann
  Publishers Inc., San Francisco, CA, USA, 1995, pp. 1137--1143.

\bibitem{10.5555/1120115.1120116}
X.~Zhu, X.~Wu, Class noise vs. attribute noise: A quantitative study of their
  impacts, Artif. Intell. Rev. 22~(3) (2004) 177–210.

\bibitem{books/wi/KaufmanR90}
L.~Kaufman, P.~J. Rousseeuw, Finding Groups in Data: An Introduction to Cluster
  Analysis., John Wiley, 1990.

\bibitem{10.5555/3001460.3001507}
M.~Ester, H.-P. Kriegel, J.~Sander, X.~Xu, A density-based algorithm for
  discovering clusters in large spatial databases with noise, in: Proceedings
  of the Second International Conference on Knowledge Discovery and Data
  Mining, KDD'96, AAAI Press, 1996, p. 226–231.

\bibitem{anonymized_project}
Exoskeleton, \href{https://exosoul.disim.univaq.it/}{The software exoskeleton}.
\newline\urlprefix\url{https://exosoul.disim.univaq.it/}

\bibitem{milne2004strategies}
G.~R. Milne, M.~J. Culnan, Strategies for reducing online privacy risks: Why
  consumers read (or don’t read) online privacy notices, Journal of
  interactive marketing 18~(3) (2004) 15--29.

\bibitem{obar2020biggest}
J.~A. Obar, A.~Oeldorf-Hirsch, The biggest lie on the internet: Ignoring the
  privacy policies and terms of service policies of social networking services,
  Information, Communication \& Society 23~(1) (2020) 128--147.

\bibitem{bhatia2016theory}
J.~Bhatia, T.~D. Breaux, J.~R. Reidenberg, T.~B. Norton, A theory of vagueness
  and privacy risk perception, in: 2016 IEEE 24th International Requirements
  Engineering Conference (RE), IEEE, 2016, pp. 26--35.

\bibitem{jensen2004privacy}
C.~Jensen, C.~Potts, Privacy policies as decision-making tools: an evaluation
  of online privacy notices, in: Proceedings of the SIGCHI conference on Human
  Factors in Computing Systems, 2004, pp. 471--478.

\bibitem{mcdonald2009comparative}
A.~M. McDonald, R.~W. Reeder, P.~G. Kelley, L.~F. Cranor, A comparative study
  of online privacy policies and formats, in: International Symposium on
  Privacy Enhancing Technologies Symposium, Springer, 2009, pp. 37--55.

\bibitem{reidenberg2015disagreeable}
J.~R. Reidenberg, T.~Breaux, L.~F. Cranor, B.~French, A.~Grannis, J.~T. Graves,
  F.~Liu, A.~McDonald, T.~B. Norton, R.~Ramanath, Disagreeable privacy
  policies: Mismatches between meaning and users' understanding, Berkeley Tech.
  LJ 30 (2015) 39.

\bibitem{mcdonald2008cost}
A.~M. McDonald, L.~F. Cranor, The cost of reading privacy policies, Isjlp 4
  (2008) 543.

\bibitem{colnago2020informing}
J.~Colnago, Y.~Feng, T.~Palanivel, S.~Pearman, M.~Ung, A.~Acquisti, L.~F.
  Cranor, N.~Sadeh, Informing the design of a personalized privacy assistant
  for the internet of things, in: Proceedings of the 2020 CHI Conference on
  Human Factors in Computing Systems, 2020, pp. 1--13.

\bibitem{hoofnagle2014alan}
C.~J. Hoofnagle, J.~M. Urban, Alan westin's privacy homo economicus, Wake
  Forest L. Rev. 49 (2014) 261.

\bibitem{urban2014privacy}
J.~M. Urban, C.~J. Hoofnagle, The privacy pragmatic as privacy vulnerable, in:
  Symposium on Usable Privacy and Security (SOUPS 2014) Workshop on Privacy
  Personas and Segmentation (PPS), 2014.

\bibitem{wilson2013privacy}
S.~Wilson, J.~Cranshaw, N.~Sadeh, A.~Acquisti, L.~F. Cranor, J.~Springfield,
  S.~Y. Jeong, A.~Balasubramanian, Privacy manipulation and acclimation in a
  location sharing application, in: Proceedings of the 2013 ACM international
  joint conference on Pervasive and ubiquitous computing, 2013, pp. 549--558.

\bibitem{brandimarte2013misplaced}
L.~Brandimarte, A.~Acquisti, G.~Loewenstein, Misplaced confidences: Privacy and
  the control paradox, Social psychological and personality science 4~(3)
  (2013) 340--347.

\bibitem{redmiles2017summary}
E.~M. Redmiles, Y.~Acar, S.~Fahl, M.~L. Mazurek, A summary of survey
  methodology best practices for security and privacy researchers, Tech. rep.
  (2017).

\bibitem{emami2020ask}
P.~Emami-Naeini, Y.~Agarwal, L.~F. Cranor, H.~Hibshi, Ask the experts: What
  should be on an iot privacy and security label?, in: 2020 IEEE Symposium on
  Security and Privacy (SP), IEEE, 2020, pp. 447--464.

\bibitem{atherton1976group}
C.~R. Atherton, Group techniques for program planning: A guide to nominal group
  and delphi processes. by andr{\'e} l. delbecq, andrew h. van de ven, and
  david h. gustafson. glenview, ill.: Scott, foresman \& co., 1975. 174 pp.
  $4.75$ paper and interpersonal conflict resolution. by alan c. filley (1976).

\bibitem{GDPR2016regulation}
T.~E. Parliament, the Council of~the European~Union, Eu general data protection
  regulation (gdpr) - regulation eu 2016/679 of the european parliament and of
  the council of 27 april 2016, Official Journal of the European Union (2016).

\end{thebibliography}
	
\end{document}